\documentclass[11pt,a4paper]{article}

\usepackage[colorlinks=true, linkcolor=black!50!blue, urlcolor=blue, citecolor=blue, anchorcolor=blue]{hyperref}
\usepackage[font=small,labelfont=bf,margin=0mm,labelsep=period,tableposition=top]{caption}
\usepackage[a4paper,top=3cm,bottom=2.5cm,left=2.5cm,right=2.5cm,bindingoffset=0mm]{geometry}

\usepackage{graphicx}
\usepackage{float}
\usepackage{afterpage}
\usepackage{epsfig,cite}
\usepackage{amssymb}
\usepackage{amsmath}
\usepackage{dsfont}
\usepackage{multirow}
\usepackage{url}
\usepackage{xcolor}
\usepackage{float}
\usepackage{afterpage}

\usepackage{url}
\usepackage{hyperref}

\usepackage{multirow,booktabs,multirow}

\newcommand{\be}{\begin{equation}}
\newcommand{\ee}{\end{equation}}
\newcommand{\bea}{\begin{eqnarray}}
\newcommand{\eea}{\end{eqnarray}}
\newcommand{\bi}{\begin{itemize}}
\newcommand{\ei}{\end{itemize}}
\newcommand{\ben}{\begin{enumerate}}
\newcommand{\een}{\end{enumerate}}
\newcommand{\la}{\left\langle}
\newcommand{\ra}{\right\rangle}

\newcommand{\lp}{\left(}
\newcommand{\rp}{\right)}

\def\frac#1#2{{{#1}\over {#2}}}
\def\gsim{\mathrel{\rlap{\lower4pt\hbox{\hskip1pt$\sim$}}
    \raise1pt\hbox{$>$}}}         
\def\lsim{\mathrel{\rlap{\lower4pt\hbox{\hskip1pt$\sim$}}
    \raise1pt\hbox{$<$}}}         

\newcommand{\draft}[1]{}

\def\beq{\begin{equation}}
\def\eeq{\end{equation}}

\def \n0{N_j^{(0)}}

\def\lapprox{\lower .7ex\hbox{$\;\stackrel{\textstyle <}{\sim}\;$}}
\def\gapprox{\lower .7ex\hbox{$\;\stackrel{\textstyle >}{\sim}\;$}}

\numberwithin{equation}{section}
\numberwithin{figure}{section}
\numberwithin{table}{section}

\usepackage{tabularx}
\newcolumntype{C}[1]{>{\centering\arraybackslash}p{#1}}

\begin{document}
\newgeometry{top=1.5cm,bottom=1.5cm,left=2.5cm,right=2.5cm,bindingoffset=0mm}
\begin{figure}[h]
  \includegraphics[width=0.32\textwidth]{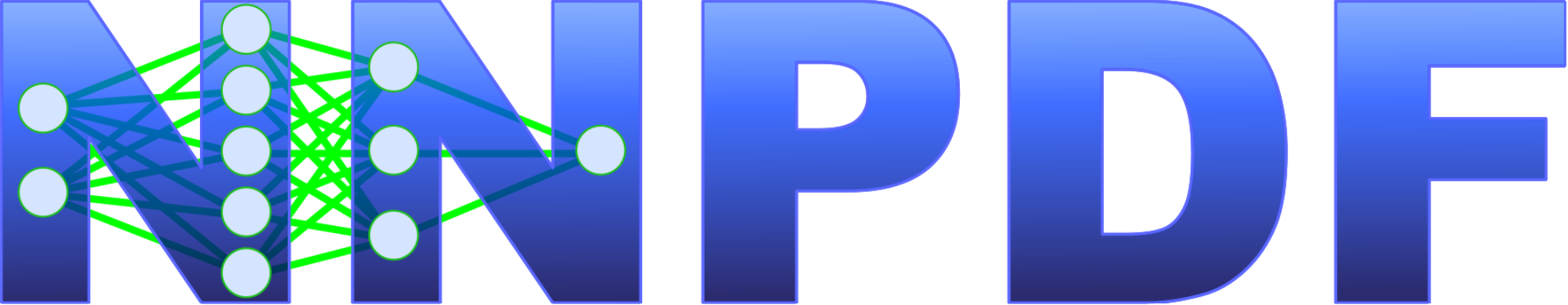}
\end{figure}
\vspace{-2.0cm}
\begin{flushright}
CERN-TH-2017-235\\
Nikhef/2017-064\\
\end{flushright}
\vspace{2cm}

\begin{center}
  {\Large \bf
    Illuminating the photon content\\[0.2cm] of the proton
    within a global PDF analysis}
\vspace{1.4cm}

  {\bf  The NNPDF Collaboration:} \\[0.1cm]
Valerio~Bertone,$^{1}$
Stefano~Carrazza,$^{2}$
Nathan~P.~Hartland,$^{1}$ and
Juan~Rojo.$^{1}$

\vspace{0.7cm}
{\it \small
~$^1$ Department of Physics and Astronomy, VU University, NL-1081 HV Amsterdam,\\
and Nikhef Theory Group, Science Park 105, 1098 XG Amsterdam, The Netherlands\\[0.1cm]
  ~$^2$ Theoretical Physics Department, CERN, CH-1211 Geneva, Switzerland\\
}

\vspace{1.0cm}

{\bf \large Abstract}

\end{center}

Precision phenomenology at the LHC requires accounting for both
higher-order QCD
and electroweak corrections as well as for photon-initiated subprocesses.
Building upon the recent NNPDF3.1 fit,
in this work the photon content of the proton is determined within a
global analysis supplemented by the LUXqed constraint relating
the photon PDF to lepton-proton
scattering structure functions:
NNPDF3.1luxQED.
The uncertainties on
the resulting photon PDF are at the level
of a few percent, with
photons carrying up to $\simeq0.5\%$ of the proton's momentum.
We study the phenomenological implications of NNPDF3.1luxQED at the LHC
for Drell-Yan, vector
boson pair, top quark pair, and Higgs plus vector boson production.
We find that photon-initiated contributions
can be significant for many processes,
leading to corrections of up to $20\%$.
Our results represent a state-of-the-art determination of the partonic structure
of the proton including its photon component.

\clearpage

\tableofcontents

\section{Introduction}
\label{sec:introduction}

Recent progress in the computation of higher-order QCD corrections to
LHC processes is such that the current state-of-the-art accuracy is
NNLO, with even N$^3$LO calculations available in some relevant cases
(see Ref.~\cite{Heinrich:2017una} for a review).
At this level of theoretical precision, the inclusion of electroweak
(EW) corrections becomes phenomenologically relevant.
With this motivation, NLO EW corrections to hard-scattering matrix
elements have been computed for many LHC processes, including single
and double vector boson, inclusive jets and dijets, and top quark pair
production, among
others~\cite{Dittmaier:2012kx,Becher:2013zua,Campbell:2016lzl,
  Czakon:2017wor,CarloniCalame:2007cd,Li:2012wna,Gavin:2012sy,Mishra:2013una,Campbell:2016dks,Kallweit:2017khh}.
Alongside progress made in process-specific calculations, the
automation of NLO EW
calculations~\cite{Frixione:2015zaa,Kallweit:2014xda,Biedermann:2017yoi}
has also advanced significantly.

In order to make the most of these developments in
the calculation of higher-order QCD and EW corrections,
equivalent progress
in the determination of the parton distribution functions (PDFs) of
the proton~\cite{Gao:2017yyd} is vital.
In this respect, most of the recent global PDF
analyses~\cite{Ball:2017nwa,Dulat:2015mca,
  Harland-Lang:2014zoa,Alekhin:2017kpj,Abramowicz:2015mha} are indeed
based on NNLO QCD theory.
On the other hand, PDF analyses that include QED and weak effects
and a determination of the photon PDF
are scarcer~\cite{Martin:2004dh,
  Giuli:2017oii,Ball:2013hta,Schmidt:2015zda}.
Such QED PDF sets are required by consistency
once EW corrections to matrix elements are included, as well
as to account for the effects of photon-initiated (PI) subprocesses.

Indeed, the inclusion of QED and weak effects into a global PDF fit
requires two main modifications.
Firstly, hard-scattering matrix
elements have to be corrected for EW effects where relevant.
This also implies taking into account the contributions from photon-induced 
subprocesses.
This leads to the second main modification, which is the
introduction of an additional parton
distribution quantifying the photon content of
the proton.
In turn, this
requires 
generalising
the DGLAP evolution equations to account for QED
corrections.
This generalisation is made possible thanks to the computation of the
splitting functions up to $\mathcal{O}\lp \alpha^2\rp$ and
$\mathcal{O}\lp
\alpha\alpha_s\rp$~\cite{deFlorian:2015ujt,deFlorian:2016gvk},
with the resulting QED-corrected evolution equations implemented in
public PDF evolution codes such as {\tt APFEL}~\cite{Bertone:2013vaa},
{\tt HOPPET}~\cite{Salam:2008qg}, and
\texttt{QEDEVOL}~\cite{Sadykov:2014aua}.

Until recently, two
distinct strategies were adopted for
the determination the photon
PDF: model calculations and data-driven approaches.
In the first case,
the photon PDF is computed on the base of a
theoretically motivated model ansatz.
The original realisation of this strategy was the MRST04QED
set~\cite{Martin:2004dh}, where the photon PDF was generated at some
low scale by one-photon collinear emission off a model for the valence quarks.
This led to a simple relation between the photon PDF and the up and
down valence distributions, which was then
evolved upwards in $Q^2$
using the DGLAP equations corrected for $\mathcal{O}(\alpha)$ contributions.

However, this  MRST04QED model accounted only
for the {\it inelastic} component of the photon PDF.
In addition to it, one should also account for the
{\it elastic} component, which can be determined
by a QED calculation in a
model-independent way~\cite{DeRujula:1998yq,Gluck:2002fi,
  Martin:2014nqa,Harland-Lang:2016qjy,
  Harland-Lang:2016apc}
in terms of the electric and magnetic form factors of the
proton.
This elastic component is derived from the equivalent photon
approximation~\cite{Budnev:1974de} and accounts for the fact that
the proton can emit photons while remaining intact.
It was furthermore shown  that the  elastic component dominates
the photon PDF at large-$x$, and that it has associated
rather small theoretical uncertainties.
In this respect, the CT14QED analysis~\cite{Schmidt:2015zda} was originally based
on the same ideas as the 
MRST04QED one,  extended with estimate of the uncertainty in their model for the
inelastic component  based on
HERA isolated photon production data~\cite{Chekanov:2009dq}, but
was subsequently complemented with an elastic
component following the procedure of~\cite{Harland-Lang:2016qjy,Harland-Lang:2016apc}.

In the second strategy, first advocated by the NNPDF
Collaboration, the photon PDF is treated on the same footing
as the quark and gluon PDFs.
Within this approach, the photon PDF is
parametrised in a model-independent way using an artificial neural
network and then
constrained by LHC Drell-Yan measurements. This procedure
was adopted in the NNPDF2.3/3.0QED
determinations~\cite{Ball:2012cx,Ball:2013hta,Ball:2014uwa,Bertone:2016ume}.
The limited sensitivity of existing LHC data to PI contributions
combined with the use of a flexible parametrisation resulted in large
uncertainties on the photon distribution.
A similar strategy was adopted in the recent analysis of
Ref.~\cite{Giuli:2017oii} in which the ATLAS 8 TeV high-mass Drell-Yan
data~\cite{Aad:2016zzw} was employed to constrain the photon
PDF.
Although this dataset is particularly sensitive to the PI
contribution, the resulting photon was still affected by large
uncertainties while a reduction in uncertainty is achieved relative to
the baseline.

Overcoming the limitations of both two strategies, the LUXqed formalism
presented in Refs.~\cite{Manohar:2016nzj,Manohar:2017eqh} represented
a breakthrough for the determination of the photon PDF. The LUXqed
methodology enhances and introduce corrections to a similar approach
adopted by earlier works in
Refs.~\cite{Anlauf:1991wr,Mukherjee:2003yh,Blumlein:1993ef}.
In this framework, both the elastic and the inelastic
components of the photon PDF can be  expressed in
terms of the electromagnetic inclusive structure functions $F_2$ and $F_L$
from lepton-proton scattering by means of an exact QED calculation.
This is very advantageous because these structure
functions are known
rather accurately both experimentally and theoretically.
Accounting for the LUXqed
 constraints then leads to a reduction of the
uncertainty of the photon PDF by more than an order of magnitude as
compared to the NNPDF3.0QED data-driven determination.

Building upon the recent NNPDF3.1
fit~\cite{Ball:2017nwa}, the goal of this paper is to perform a
global PDF analysis including QED corrections where the LUXqed
calculation is used to constrain the photon PDF.
The resulting PDF set, NNPDF3.1luxQED,
represents a state-of-the-art determination of the
partonic content of the proton including its photon
component.
The uncertainties on the photon PDF are now at the level
of a few percent, with
photons carrying up to $\simeq0.5\%$ of the total proton's
momentum.\footnote{See also~\cite{Harland-Lang:2017dzr} for  related studies in the MMHT framework.}
Comparing with NNPDF3.0QED, we find good
  agreement within uncertainties in the $x\gsim 0.02$ region, and
larger differences for smaller values of $x$.

We also take a first look at
the phenomenological implications
of NNPDF3.1luxQED for photon-initiated
processes at the LHC.
Previous studies based on NNPDF2.3/3.0QED
indicated that PI contributions were potentially large, particularly at large
invariant masses or transverse momenta, for processes such as
Drell-Yan, $W$ pair, and top-quark pair
production~\cite{Bertone:2015lqa,
  Mangano:2016jyj,Bourilkov:2016qum,Accomando:2016tah,
  Pagani:2016caq,Czakon:2017mmr}.
Indeed, PI effects
represented in some cases the dominant source of
theoretical uncertainty.
We find that photon-initiated corrections
computed with NNPDF3.1luxQED
can be significant for many processes,
leading to corrections of up to $20\%$ depending
  on the kinematics.
These PI contributions
are consistent with previous estimates
based on NNPDF3.0QED within uncertainties in the
  kinematic region $Q \gsim M_Z$, with larger
differences in processes for which $Q < M_Z$.

The outline of this paper is as follows.
In Sect.~\ref{sec:fitsettings} we present the settings of the  global
NNPDF3.1luxQED analysis in terms of input experimental data, theoretical
calculations, and fitting strategy.
Then in Sect.~\ref{sec:results} we present the NNPDF3.1luxQED set,
including a discussion of the momentum fraction of the proton carried
by the photon, and in Sect.~\ref{sec:pheno} we discuss some of its
phenomenological implications for PI processes at the LHC.
In Sect.~\ref{sec:conclusion} we summarise and discuss how our results
and the code used to produce them are made publicly available.
The full breakdown of the $\chi^2/N_{\rm dat}$ values in
NNPDF3.1luxQED and its comparison with those in NNPDF3.1 is collected in
Appendix~\ref{sec:appendix}.

\section{Fit settings}\label{sec:fitsettings}

In this section we describe the fit configuration of the
NNPDF3.1luxQED global analysis.
We begin with a review of the input experimental dataset.
We then discuss the theoretical framework, including a
short summary of the relevant aspects of the LUXqed formalism
along with the treatment of QED effects in the DGLAP evolution and the
DIS structure functions.
Finally, we present the strategy adopted to include the photon PDF in
the global fit accounting for the LUXqed theoretical
constraints.

\subsection{Experimental data}

The NNPDF3.1luxQED analysis is based on the same dataset as the recent NNPDF3.1
global fit~\cite{Ball:2017nwa}.
This dataset includes fixed-target~\cite{Arneodo:1996kd,Arneodo:1996qe,bcdms1,
bcdms2,Whitlow:1991uw,Onengut:2005kv,Goncharov:2001qe,MasonPhD} and HERA~\cite{Abramowicz:2015mha}
inclusive DIS measurements; charm and bottom cross-sections from HERA~\cite{Abramowicz:1900rp};
fixed-target Drell-Yan production~\cite{Webb:2003ps,Webb:2003bj,Towell:2001nh,Moreno:1990sf}; Tevatron
gauge boson and inclusive jet production~\cite{Aaltonen:2010zza,Abazov:2007jy,
Aaltonen:2008eq,Abazov:2013rja,D0:2014kma}; along with electroweak boson production,
inclusive jet, and $t\bar{t}$ cross-sections from ATLAS~\cite{Aad:2011dm,Aad:2013iua,
Aad:2011fp,Aad:2011fc,Aad:2013lpa,ATLAS:2012aa,ATLAS:2011xha,
TheATLAScollaboration:2013dja,Aad:2015auj,Aaboud:2016btc,Aad:2014kva,Aaboud:2016pbd,
Aad:2015mbv,Aad:2014qja,Aad:2014xaa}, CMS~\cite{Chatrchyan:2012xt,Chatrchyan:2013mza,
Chatrchyan:2013tia,Chatrchyan:2013uja,Chatrchyan:2013faa,Chatrchyan:2012bra,
Chatrchyan:2012ria,Khachatryan:2016pev,Khachatryan:2015luy,Khachatryan:2016mqs,
Khachatryan:2015oqa,Khachatryan:2015oaa} and LHCb~\cite{Aaij:2012vn,Aaij:2012mda,
Chatrchyan:2012bja,Aaij:2015gna,Aaij:2015zlq}. We refer to
Ref.~\cite{Ball:2017nwa} for details about the implementation of each
experiment. 

For consistency, in this study we use exactly the same dataset as in
NNPDF3.1, and in particular the same choice of kinematic cuts.
Note that a number of those cuts were determined with the aim
of minimising the potential effects from EW corrections and PI contributions.
This choice implies that the kinematic regions more sensitive to PI
effects are deliberately cut away.
In addition, we do not include some recent measurements with
known sensitivity to the photon PDF, such as the
ATLAS high-mass Drell-Yan measurement at 8 TeV~\cite{Aad:2016zzw}, since these
were not part of the NNPDF3.1 dataset.

\subsection{The LUXqed formalism}
\label{sec:theory}

We briefly review the LUXqed formalism for the determination of
the photon PDF, focusing on those features relevant to its
implementation in a global analysis.
For a comprehensive discussion we refer the reader to
Refs.~\cite{Manohar:2016nzj,Manohar:2017eqh}.
In the LUXqed procedure, the photon PDF can be expressed in
terms of the lepton-proton scattering inclusive structure functions $F_2$ and
$F_L$ by means of an exact QED calculation as follows:
\begin{equation}\label{eq:luxqed}
\begin{array}{l}
\displaystyle x\gamma(x,\mu)=\frac{1}{2\pi\alpha(\mu)}\int_x^1
\frac{dz}{z}\Bigg\{ \int_{Q^2_{\rm \min}}^{\mu^2/(1-z)}
\frac{dQ^2}{Q^2}\alpha^2(Q^2)\Bigg[ -z^2 F_L(x/z,Q^2) \\
\\
\displaystyle+\lp zP_{\gamma q}(z)+\frac{2x^2m_p^2}{Q^2}\rp F_2(x/z,Q^2)\Bigg] -\alpha^2(\mu)
z^2F_2(x/z,\mu^2)\Bigg\} + \mathcal{O}\lp\alpha\alpha_s,\alpha^2\rp,
\end{array}
\end{equation}
where $m_p$ is the proton mass, $\mu$ is the factorisation scale, $x$
and $z$ are the momentum fractions, $\alpha$ the running QED coupling,
and $P_{\gamma q}$ the photon-quark splitting function.
The lower integration limit in the $Q^2$ integral is given by
$Q^2_{\rm \min}=(m_p^2x^2)/(1-z)$.

Note that the integral in $z$ in Eq.~(\ref{eq:luxqed}) extends up to
$z=1$. Therefore the LUXqed photon has an explicit dependence
upon the elastic component of the structure functions, proportional to
$\delta(1-z)$.
This component can be expressed in terms of the electric and magnetic
form factors $G_E$ and $G_M$.
In~\cite{Manohar:2016nzj,Manohar:2017eqh}, the elastic component of
$\gamma(x,\mu)$ is determined using the form factors extracted from a
fit to world data by the A1 collaboration~\cite{Bernauer:2013tpr} for
$Q^2 \le 10$ GeV$^2$. The dipole model is then used to extrapolate the
form factors to larger values of $Q^2$.
A corresponding uncertainty due to the treatment of the large-$Q^2$
extrapolation region is included in the evaluation of the photon PDF.

Furthermore, we observe that the $Q^2$ integral in
Eq.~(\ref{eq:luxqed}) requires an understanding of the structure
functions down to potentially very low scales, well outside the region
where perturbative QCD is applicable.
Following the prescription of
Refs.~\cite{Manohar:2016nzj,Manohar:2017eqh}, the integration in $Q^2$
of the inelastic component ($z<1$) of $F_2$ and $F_L$ is achieved by
combining parameterisations of experimental data with the perturbative
computation in terms of PDFs where appropriate.
Specifically, contributions to the inelastic structure functions come from two
regions separated by $W^2= m_p^2+Q^2(1-z)/z$.
In the resonance region, defined
$(m_p+m_{\pi})^2 \le W^2\le 3.5$ GeV$^2$, the fit of the CLAS
collaboration is used~\cite{Osipenko:2003bu}. In order to assess the
uncertainty due to this choice, the parametrisation of
Ref.~\cite{Christy:2007ve} is also considered.

The continuum region, defined as $W^2 \geq 3.5$ GeV$^2$, is further
subdivided into two regions according the value of $Q^2$. For $Q^2 \le Q^2_{\rm
match}$, with $Q^2_{\rm match} = 9$ GeV$^2$, the GD11-P fit by
HERMES~\cite{Airapetian:2011nu,Abramowicz:1991xz} is employed. For $Q^2 >
Q^2_{\rm match}$, structure functions are computed in terms of PDFs by
means of their factorised expressions. The value of $Q^2_{\rm match}$ can be
varied to estimate the uncertainty associated with this particular choice.

As a part of the present work, the LUXqed formalism has been
implemented in an open-source public library, {\tt
  FiatLux}~\cite{stefano_carrazza_2017_1117325}, which has been used
to produce the NNPDF3.1luxQED fits.
The results obtained with {\tt FiatLux} have been benchmarked with the
original implementation used to produce the results of
Refs.~\cite{Manohar:2016nzj,Manohar:2017eqh}, finding excellent
agreement.

\subsection{Theoretical calculations}
\label{sec:theocalculations}

The QCD calculations used in the present analysis are identical to those used in
NNPDF3.1~\cite{Ball:2017nwa}.
DGLAP evolution and DIS structure functions are computed at NLO and NNLO
accuracy in QCD using {\tt APFEL}~\cite{Bertone:2013vaa}. Heavy-quark mass
effects in the structure functions are included using the FONLL general-mass
variable-flavour-number scheme~\cite{Forte:2010ta}, and the charm PDF is fitted
to data on an equal footing as the light quark and gluon
PDFs~\cite{Ball:2015tna,Ball:2015dpa,Ball:2016neh}.
Hadronic observables are computed at NLO using fast interpolation
tables in the {\tt APPLgrid}~\cite{Carli:2010rw} and {\tt
  fastNLO}~\cite{Wobisch:2011ij} formats and combined to the DGLAP
evolution kernels using {\tt APFELgrid}~\cite{Bertone:2016lga}.
NNLO corrections to hadronic processes are included by means of
point-by-point NNLO/NLO $K$-factors.

In the NNPDF3.1luxQED fit, QCD corrections are supplemented with
QED effects.
Concerning the evolution of PDFs, on top of the $\mathcal{O}(\alpha)$
corrections, also the $\mathcal{O}\lp \alpha^2\rp$ and
$\mathcal{O}\lp \alpha\alpha_s\rp$ splitting functions computed in
Refs.~\cite{deFlorian:2015ujt,deFlorian:2016gvk} are included in the
DGLAP evolution equations.
Additionally, $\mathcal{O}(\alpha)$ corrections to the DIS coefficient
functions are included. This introduces an additional (but mild)
sensitivity to the photon PDF.
The implementation in {\tt APFEL} of the aforementioned QED
corrections to the DIS structure functions and to the DGLAP evolution
equations, together with the corresponding benchmarking, were presented in
Ref.~\cite{Giuli:2017oii}.
On the other hand, as in NNPDF3.1, pure weak corrections to hadronic
observables are not accounted for.

\begin{figure}[h]
\begin{center}
  \includegraphics[scale=0.45]{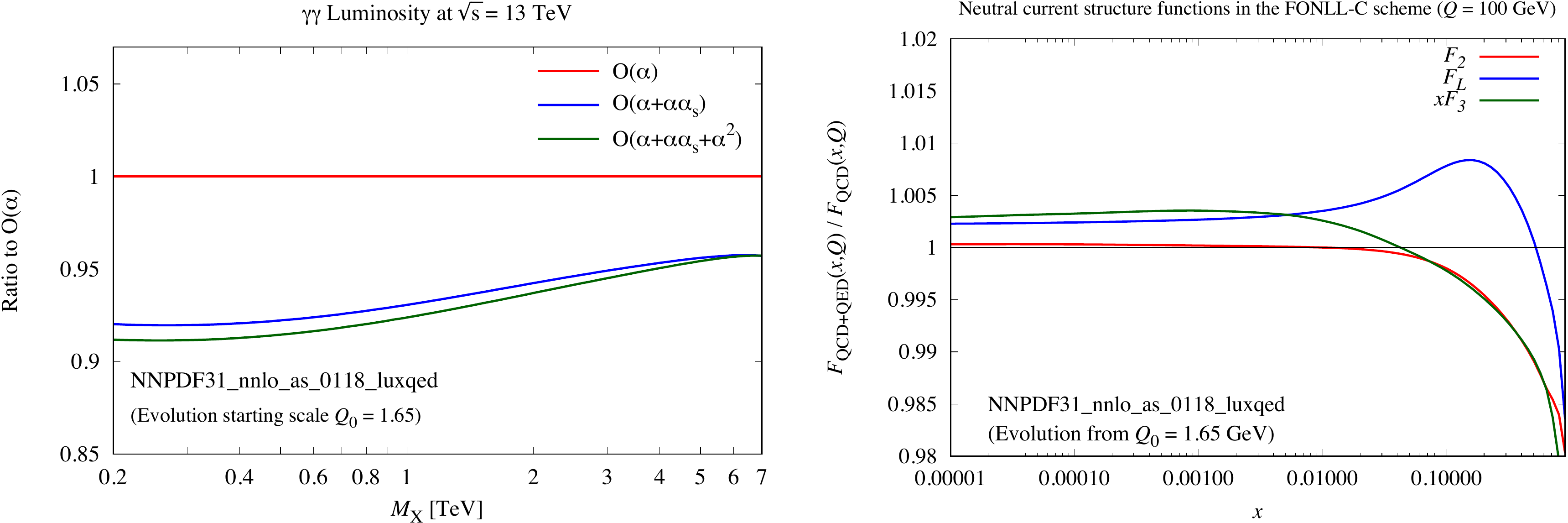}
\end{center}
\vspace{-0.5cm}
\caption{\small Left: comparison of the $\mathcal{L}_{\gamma\gamma}$
  luminosity at $\sqrt{s}=13$ TeV,
  computed starting from NNPDF3.1luxQED 
  at $Q_0=1.65$ GeV and then evolving upwards using
  different types of QED corrections in the DGLAP
  splitting functions.
  Right: comparison between the DIS splitting functions $F_2$,
  $F_L$, and $xF_3$ at $Q=100$ GeV with and without
  QED effects included. \label{fig:lumisQED}
}
\end{figure}

Using NNPDF3.1luxQED as an input, we find that the cumulative effect
on the photon-photon luminosity $\mathcal{L}_{\gamma\gamma}$ of the
$\mathcal{O}\lp \alpha^2\rp$ and $\mathcal{O}\lp \alpha\alpha_s\rp$
corrections to the DGLAP splitting functions ranges between
$\simeq 10\%$ at low invariant masses $M_X$ and $\simeq 5\%$ for high
$M_X$, see Fig.~\ref{fig:lumisQED}.
Since these effects are larger than the typical uncertainties on the
photon PDF determined through the LUXqed
approach, it is important to
take them into account.
Concerning the DIS structure functions, using NNPDF3.1luxQED,
as shown in Fig.~\ref{fig:lumisQED} one
finds that the overall impact of the QED effects is at the permille level,
except at large values of $x$ where they can be up to 2\%.

In data-driven determinations of the photon PDF,
it is in general necessary to
include constraints from observables sensitive to PI processes.
For instance, in the analysis of Ref.~\cite{Giuli:2017oii} PI
contributions to the ATLAS 8 TeV high-mass Drell-Yan
data~\cite{Aad:2016zzw} were computed via the {\tt aMCfast}
interface~\cite{aMCfast} to the Monte Carlo generator {\tt
  MadGraph5\_aMC@NLO}~\cite{Alwall:2014hca} and used to constrain the
photon PDF.
In the present study, the photon PDF is determined
from a global analysis of hard-scattering data supplemented
by the LUXqed theoretical constraint of Eq.~(\ref{eq:luxqed}).
Given that the NNPDF3.1 dataset and the associated kinematic cuts were
specifically designed to minimise the effects of PI contributions, one
expects the impact of PI processes to hadronic cross-sections in
NNPDF3.1luxQED to be minimal, and thus they are not included here.
We have explicitly verified for some of the NNPDF3.1
datasets that this is an excellent
approximation, see also the
comparisons of Sect.~\ref{sec:pheno}.
Nevertheless, the approach presented in this work is fully general, and future NNPDF analyses
with QED corrections will include collider measurements characterised
by sizeable PI contributions.

\subsection{Fitting strategy}
\label{sec:fittingstrategy}

The determination of the NNPDF3.1luxQED set is performed by means of
an iterative procedure.
The starting point is a prior set of quark and gluon PDFs, in this
case NNPDF3.1.
From this PDF set, the high-$Q^2$ inelastic component of the photon
PDF is computed at $Q=100$ GeV using Eq.~(\ref{eq:luxqed}), while for
the other components the same inputs as in Ref.~\cite{Manohar:2017eqh}
are adopted.
The resulting photon PDF is then evolved down to the
parametrisation scale $Q_0=1.65$ GeV and
used as a fixed input in a refit of quark and gluon PDFs.

\begin{figure}[t]
\begin{center}
  \includegraphics[scale=0.30]{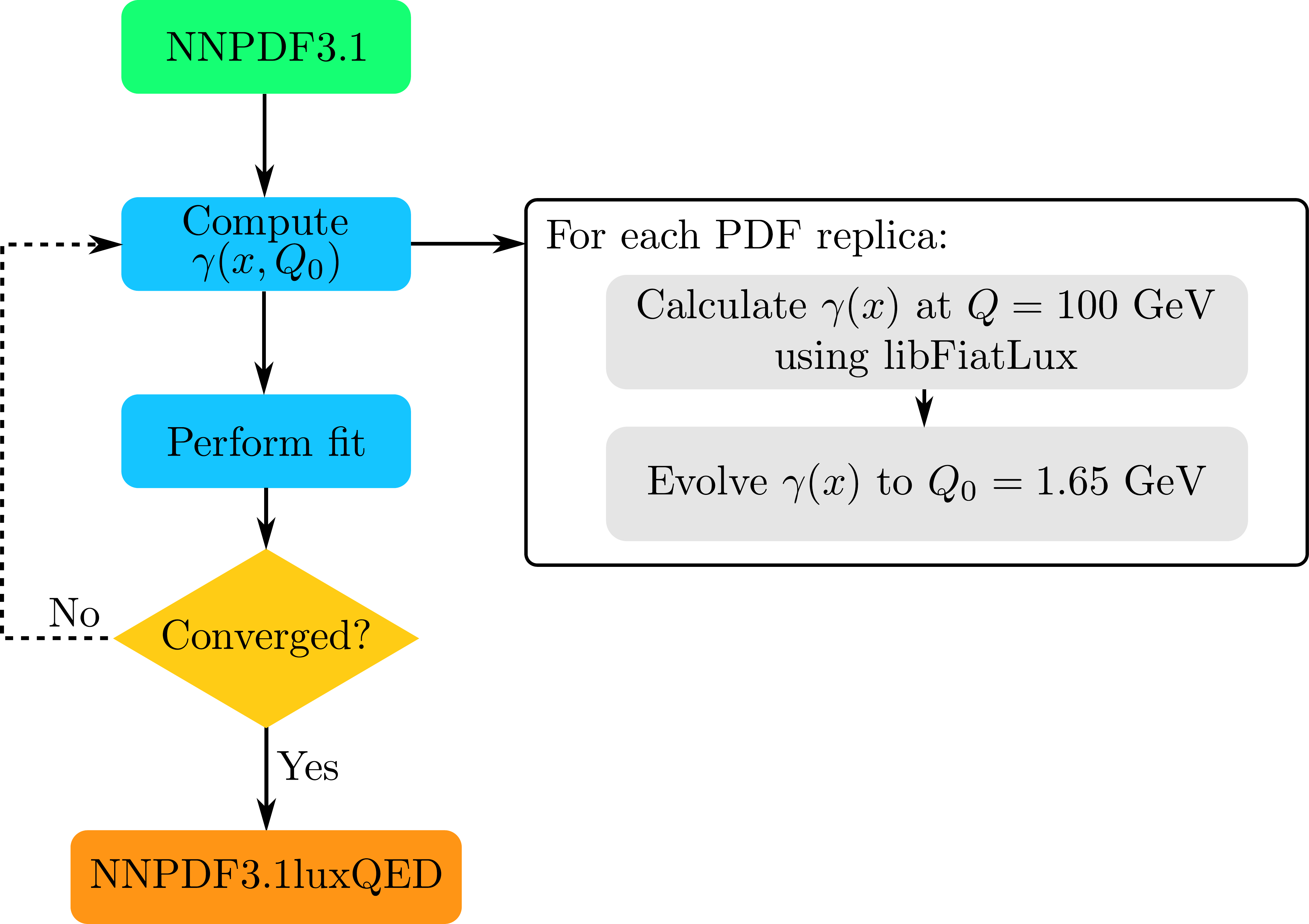}
  \caption{\small Flow diagram representing the NNPDF3.1luxQED fitting
    strategy.
    In the last iteration $n_{\rm ite}$, once the procedure has converged,
    the additional LUXqed17 are added to $\gamma(x,Q)$, see
    Sect.~\ref{sec:extratherrors}.
   }\label{fig:fittingstrategy} 
  \end{center}
\end{figure}

In this refit, the DGLAP evolution equations consistently include QED
effects, the PI contribution to the DIS structure functions is taken
into account, and the momentum sum rule reads
\begin{equation}\label{sec:momentumsumrule}
  \int_0^1 dx\,x\lp
  \Sigma(x,Q_0)+g(x,Q_0)+\gamma(x,Q_0)\rp=1\,.
\end{equation}
This procedure, schematically illustrated in
Fig.~\ref{fig:fittingstrategy}, is repeated until convergence is
reached, in the sense that a stable photon PDF, and thus stable quark
and gluon PDFs, are obtained.
We consider stable the results of two consecutive iterations where the
central value of the photon PDF varies by less than 5\% of its
uncertainty.

As usual in the NNPDF approach, PDF uncertainties are represented by
means of an ensemble of $N_{\rm rep}$ Monte Carlo replicas. Each
replica is required to meet a set of quality criteria, discussed in
Ref.~\cite{Ball:2014uwa}, with fits failing these criteria being
discarded.
As the present study involves an iterative procedure, one must start
with a sample of replicas large enough such that once all
$n_{\rm ite}$ iterations have been completed a significant number of
replicas still survives.
To this end, here we use a prior sample of $N_{\rm rep}=500$ replicas.

Each replica will lead to different high-$Q^2$ DIS structure functions
and therefore, by virtue of Eq.~(\ref{eq:luxqed}), to a different
photon PDF that is used as an external constraint in the following
fit.
The resulting quark and gluon PDFs are then used as an
input to the following iteration of the fitting procedure, until convergence
is reached.
As the NNPDF3.1 dataset is (by construction) relatively
insensitive to the photon PDF, the convergence is rapid and
results are stable already after the second iteration.
In future analyses, when hadronic measurements sensitive to PI
contributions will be included, convergence is likely to be slower.

There are two main differences between our strategy and the
direct application of Eq.~(\ref{eq:luxqed}) to NNPDF3.1.
Firstly, the influence of the photon PDF in the DGLAP evolution
equations and in the DIS structure functions is consistently taken
into account during the fit of the quark and gluon PDFs.
Secondly, the contribution of the photon PDF to the total momentum
fraction is properly treated by imposing the momentum sum rule
Eq.~(\ref{sec:momentumsumrule}) during the fits.
While these effects are likely to be small in this specific analysis,
our framework is fully general and allows for the consistent inclusion
of hadronic observables sensitive to the photon-initiated contributions.

\begin{figure}[t] \begin{center}
    \includegraphics[scale=0.50]{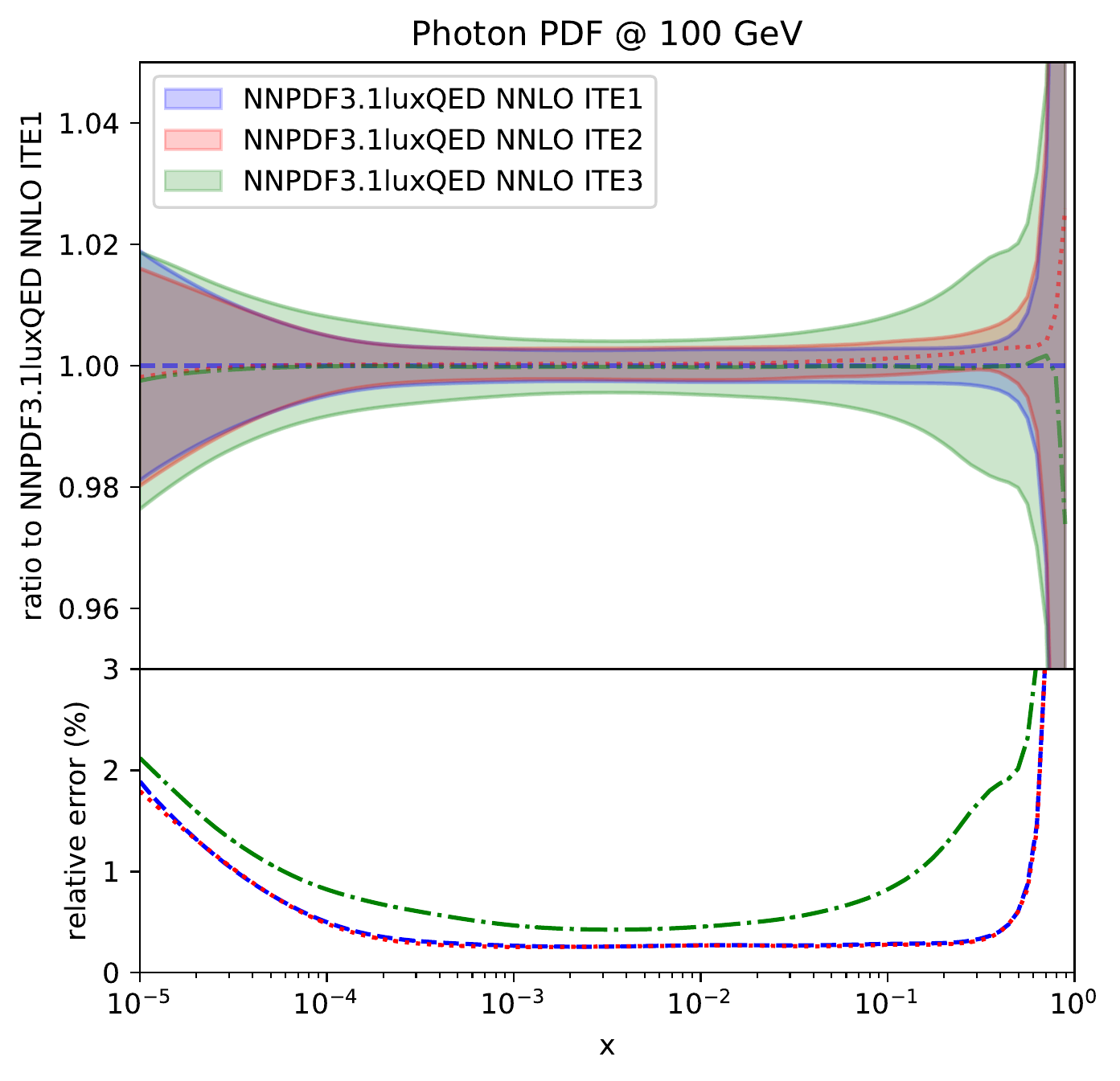}\includegraphics[scale=0.50]{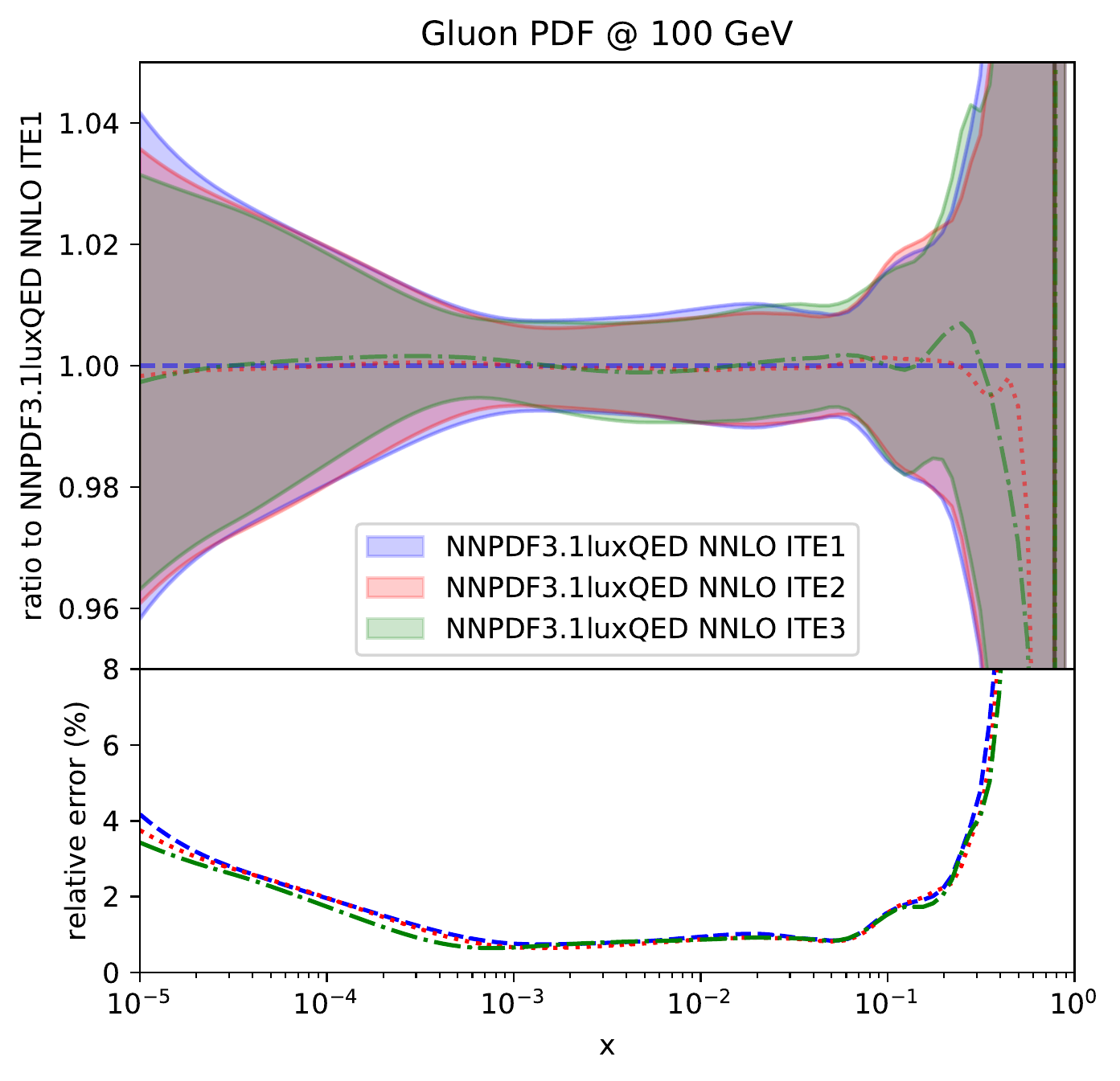}
    \includegraphics[scale=0.50]{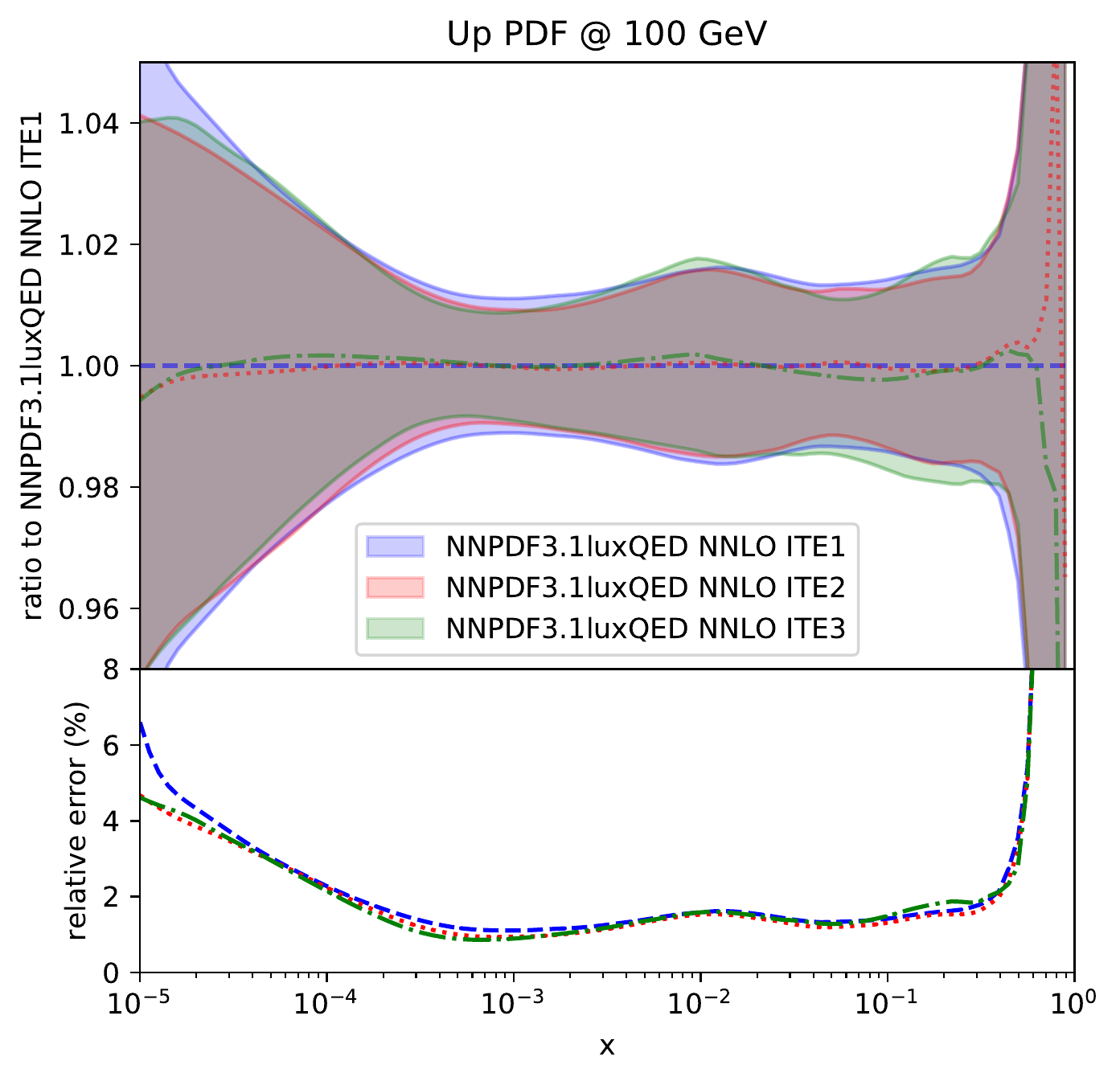}\includegraphics[scale=0.50]{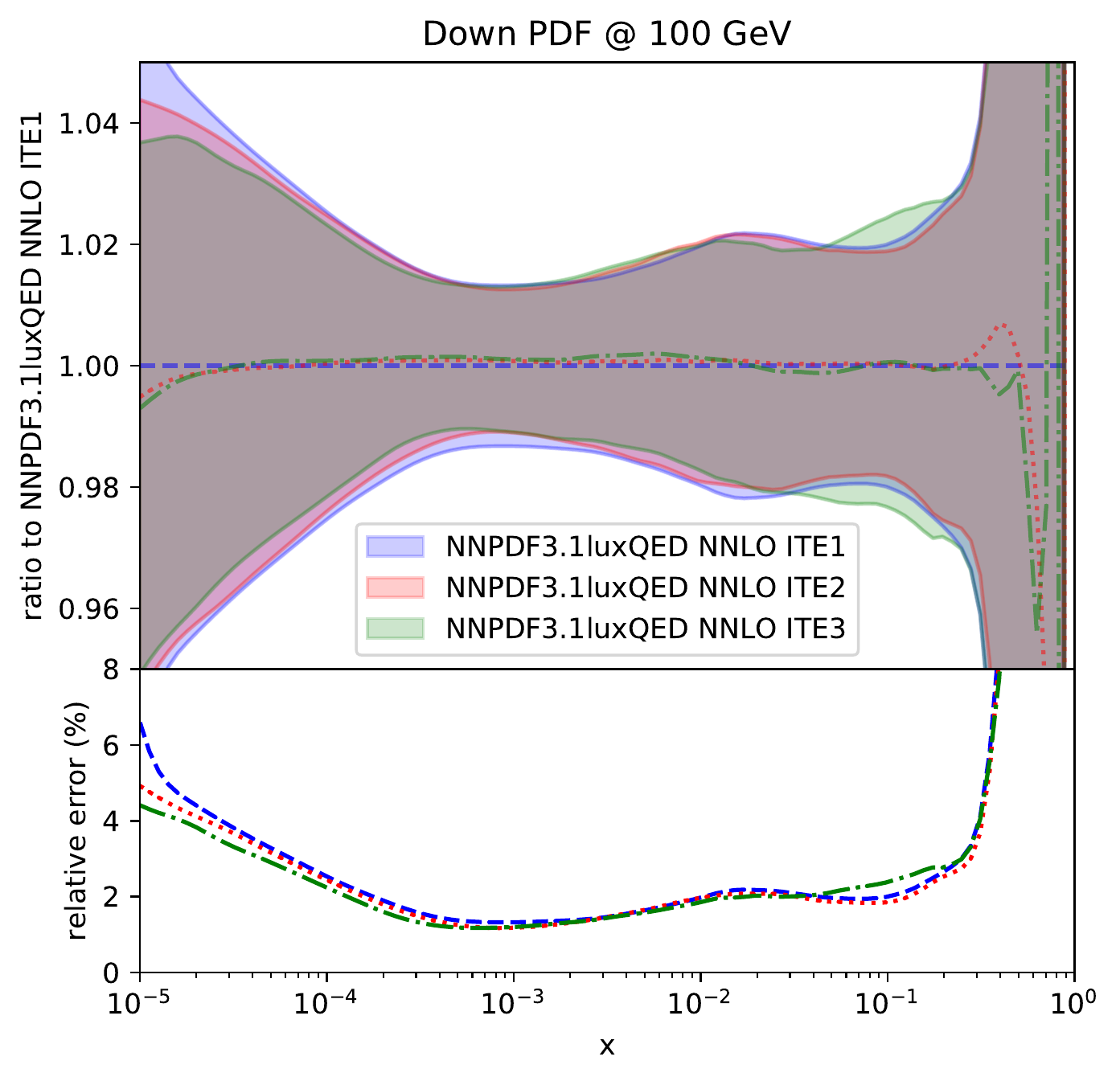}
    \caption{\small Comparison of the photon, gluon, up quark, and down quark
      PDFs at $Q=100$ GeV between the first and second iterations fit
      (ITE1 and ITE2 respectively) of the fitting procedure
      sketched in Fig.~\ref{fig:fittingstrategy}.
      For completeness, we also show the third and final iteration
      of the procedure $n_{\rm ite}$
      (ITE3) where the additional
      LUXqed17 systematic variations have been
      added to $\gamma(x,Q)$, see
      Sect.~\ref{sec:extratherrors} \label{fig:convergenceiterations} }
\end{center} \end{figure}

In order to illustrate the convergence of the procedure, in
Fig.~\ref{fig:convergenceiterations} we show a comparison of the
photon, gluon, up quark, and down quark PDFs at $Q=100$ GeV between
the first and the second iteration (labelled ITE1 and ITE2, respectively).
We have verified that additional iterations leave the photon PDF
unchanged, demonstrating that stability has been reached.
For completeness, in Fig.~\ref{fig:convergenceiterations} we also show
the third and final iteration of the procedure (ITE3), where the
additional LUXqed17 systematic variations are added to the photon PDF
(see Sect.~\ref{sec:extratherrors}).
As expected, these have the largest impact in the region $x\gsim 0.05$, where the
elastic contribution to the photon PDF is most important.
      
\subsection{The uncertainties on the photon PDF}
\label{sec:extratherrors} 

As mentioned above, the calculation of the photon PDF in terms of
structure functions involves several contributions: the elastic
component, the inelastic resonance component, and the inelastic low-
and high-$Q^2$ continuum components.
Only the last component can be factorised in terms of PDFs and
perturbative coefficient functions.
Therefore, the ensemble of $N_{\rm rep}$ Monte Carlo replicas of the
photon PDF accounts only for a part of the uncertainty, namely the one
associated to the inelastic high-$Q^2$ component.
For a comprehensive estimate of the uncertainty one must also account
for a number of additional sources of error.

The following sources of uncertainty are
considered~\cite{Manohar:2017eqh}: the elastic contribution from the
A1 world proton form factor fits~\cite{Bernauer:2013tpr}; the
parametrisation of the DIS structure functions in the resonance
region~\cite{Christy:2007ve,Osipenko:2003bu,Airapetian:2011nu}; the
parametrisation of $R_{\rm L/T}$~\cite{Abramowicz:1991xz,Liang:2004tj,
  Abe:1998ym}, the ratio between longitudinal and transverse structure
functions; the scale $Q_{\rm match}^2$ at which low- and high-$Q^2$
inelastic structure functions are matched; a twist-4 modification of
the longitudinal structure function
$F_L$~\cite{Cooper-Sarkar:2016foi,Harland-Lang:2016yfn}; and finally
an estimate of the missing higher-order corrections in the calculation
of the DIS structure functions at high $Q^2$.

In the NNPDF3.1luxQED analysis, these uncertainties are introduced at
the last iteration of the procedure.
Once the quark and gluon PDFs from the $(n_{\rm ite}-1)$-th iteration have
been determined, they are used to construct
$\gamma_{n_{\rm ite}}(x,Q)$.
Then, for each photon PDF replica of this last iteration,
$\gamma_{n_{\rm ite}}^{(k)}$, $n_{\rm sys}=7$ extra uncertainties are
included as statistical fluctuations upon the photon PDF at $Q=100$
GeV with correlations in $x$, namely:
\begin{equation}
\label{eq:ite3syst}
\widetilde{\gamma}_{n_{\rm ite}}^{(k)}(x,Q)
=
\gamma_{n_{\rm ite}}^{(k)}(x,Q)+\sum_{j=1}^{n_{\rm sys}}
\delta\gamma_j^{\rm (lux)}(x,Q)\cdot
\mathcal{N}(0,1) \, , \quad k=1,\ldots,N_{\rm rep} \, ,
\end{equation}
where $\mathcal{N}(0,1)$ is an univariate Gaussian random number and
$\delta\gamma_j^{\rm (lux)}$ is the normalised eigenvector for the
$j$-th systematic uncertainty in LUXqed17.
Specifically, $\delta\gamma_j^{\rm (lux)}$ is obtained through the diagonalisation
of the  covariance matrix for the extra LUXqed17 uncertainties defined on a grid of $x$
points, using a similar method as that of Refs.~\cite{Carrazza:2015aoa,Carrazza:2016htc}.
We have verified that this approach is numerically equivalent to
using the corresponding Hessian eigenvectors of LUXqed17.

The photon PDF defined in Eq.~(\ref{eq:ite3syst}) is finally used as
an input for the $n_{\rm ite}$-th fit iteration, to determine the final
set of quark and gluons of NNPDF3.1luxQED.

\section{The NNPDF3.1luxQED set}
\label{sec:results}

In this section we discuss the results of the iterative procedure
outlined in Sect.~\ref{sec:fittingstrategy}: the NNPDF3.1luxQED fit.
We focus on the NNLO case and comment where appropriate on any
differences with respect to NLO.
The overall fit quality of NNPDF3.1luxQED is
$\chi^2/N_{\rm dat} = 1.168$ at NLO and $\chi^2/N_{\rm dat} = 1.148$
at NNLO.
While there is some variation dataset-to-dataset, the global fit
quality is identical to the corresponding NNPDF3.1 results at NNLO.
See Appendix~\ref{sec:appendix} for a full breakdown of the data
description at NNLO and its comparison with NNPDF3.1.

\subsection{The photon PDF}
\label{sec:PDFcomparisonPhoton}

\begin{figure}[t]
  \begin{center}
    \includegraphics[scale=0.55]{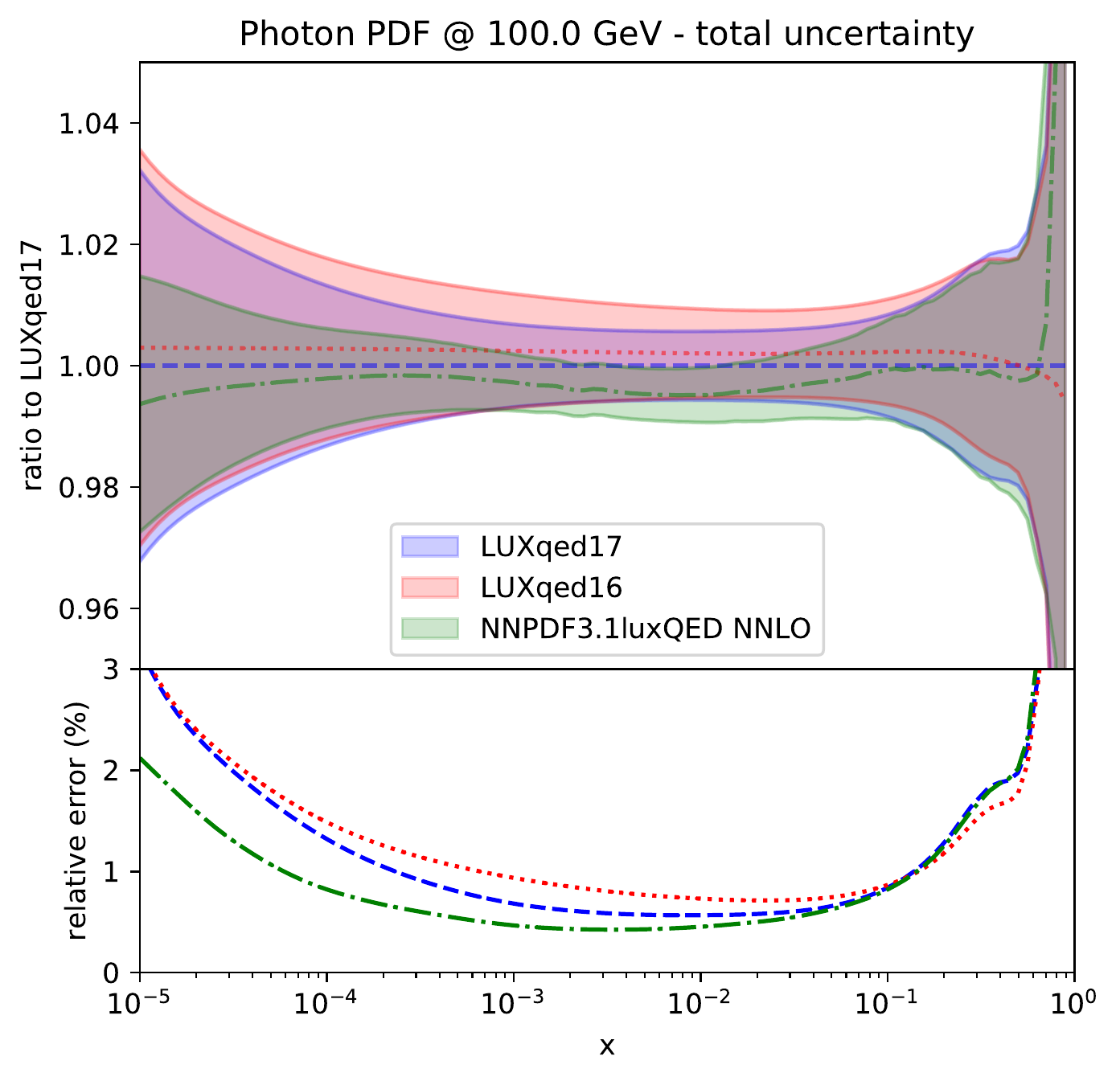}
    \includegraphics[scale=0.55]{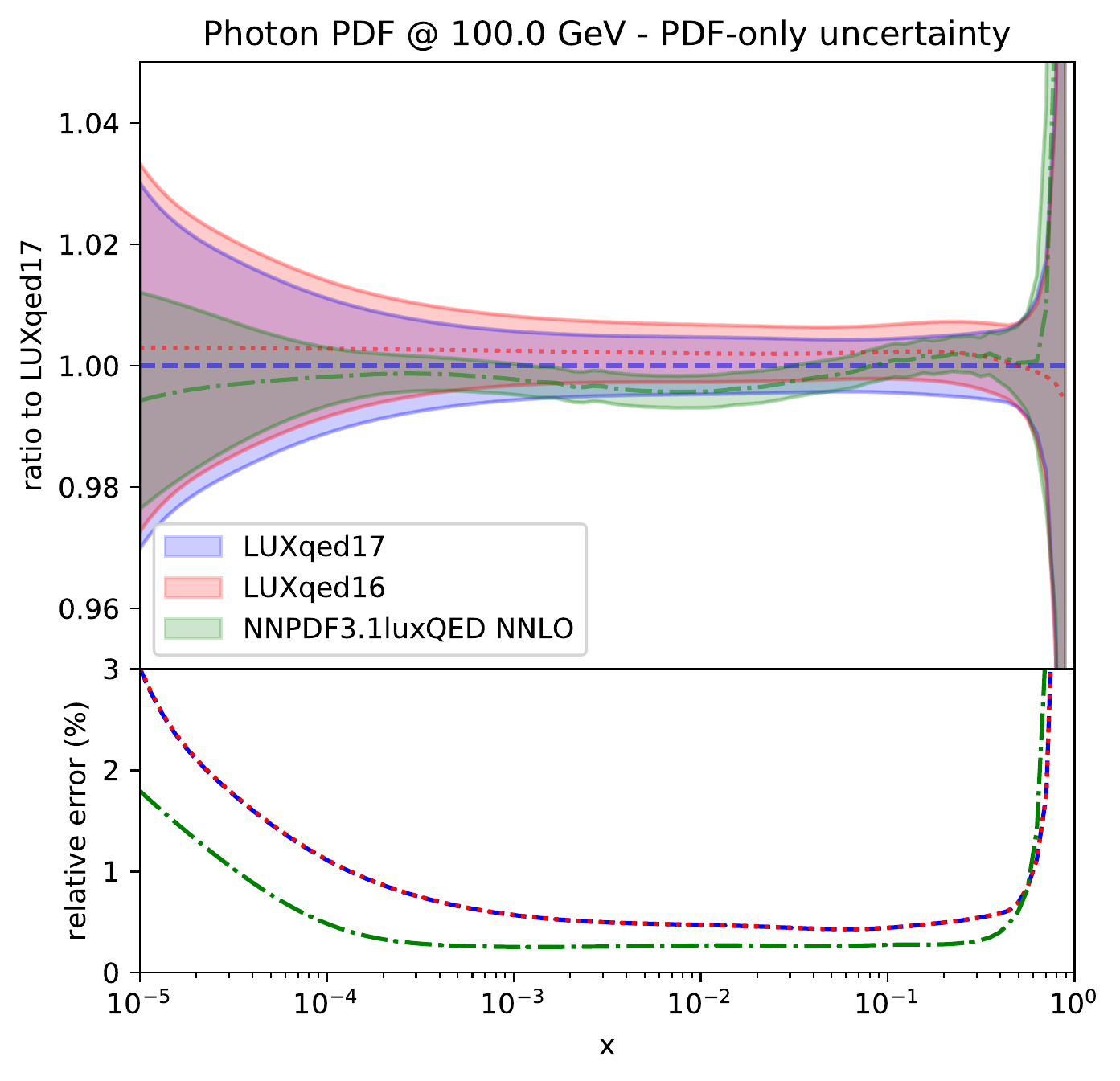} \caption{\small Left:
        comparison of the NNPDF3.1luxQED photon at $Q=100$ GeV with that
        of LUXqed16/17 normalized to
        the central value of the latter.
        The bottom panel indicates the relative uncertainty
        on the photon PDF in each case.
    Right: the same comparison, now including only the uncertainties on
    $\gamma(x,Q)$ related to the quark and gluon PDFs
    in the high-$Q^2$ inelastic component.
    \label{fig:photoncomp1} } \end{center} \end{figure}

Here we compare our results for the photon PDF $\gamma(x,Q)$ with
those of the NNPDF3.0QED and LUXqed16/17 PDF sets.
Comparisons with the latter are performed always at $Q \ge 10$ GeV, as
the LUXqed16/17 sets are not defined below this scale.
As discussed in Sect.~9.2 of Ref.~\cite{Manohar:2017eqh},
the LUXqed17 set has a improved evaluation of the photon PDF
calculation and of the associated error estimates
in comparison to LUXqed16.
In the left panel of Fig.~\ref{fig:photoncomp1} we compare the
NNPDF3.1luxQED photon PDF at $Q=100$ GeV with the corresponding
results from LUXqed16 and LUXqed17, normalised to the central value of
the latter.
The three determinations agree well across the full $x$ range, with
central values always compatible within uncertainties.
In addition, for $x\gsim 0.1$ the total uncertainties on the photon
PDF from NNPDF3.1luxQED and LUXqed16/17 are identical.
This feature is explained by the fact that in this region the
uncertainties due to the elastic and low-$Q^2$ inelastic structure
functions dominate.

At medium- and small-$x$, the NNPDF3.1luxQED photon exhibits
somewhat smaller uncertainties. This is due to the use of a different
set of quark and gluon PDFs determining the high-$Q^2$
inelastic component, specifically NNPDF3.1 rather than the PDF4LHC15 set~\cite{Butterworth:2015oua}
used in LUXqed16/17.
The contribution from the different error sources is further
illustrated in the right panel of Fig.~\ref{fig:photoncomp1} where the
same comparison including only the uncertainties due to the high-$Q^2$
inelastic component is shown.
The plot shows how at medium- to small-$x$ the contribution from the
high-$Q^2$ inelastic structure functions dominates the overall
uncertainty.

\begin{figure}[t]
\begin{center}
  \includegraphics[scale=0.50]{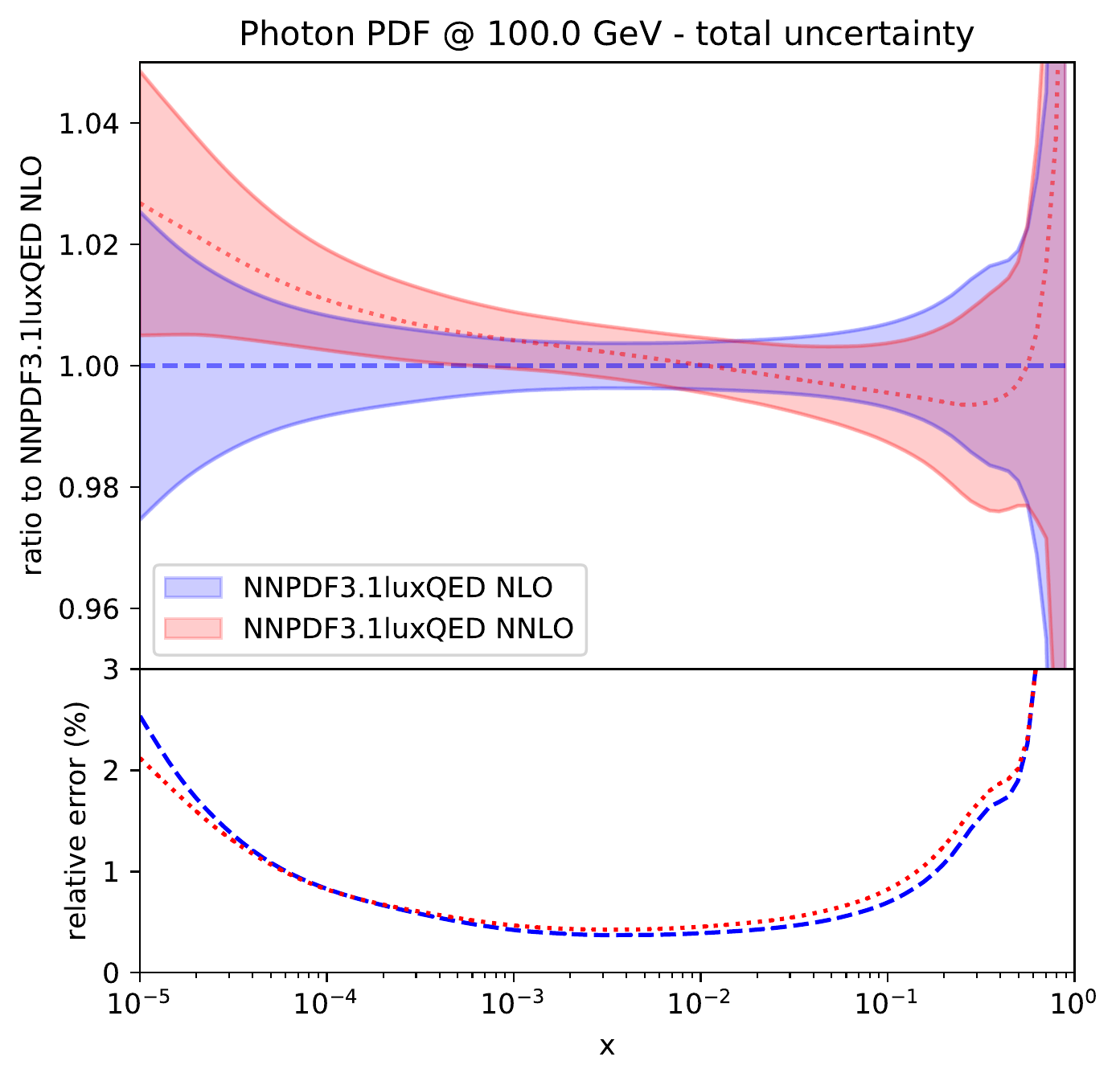}
  \caption{\small
    Comparison between $\gamma(x,Q)$ in the NNPDF3.1luxQED NLO and NNLO fits.
    \label{fig:photoncomp2}
  }
\end{center}
\end{figure}

In order to gauge the stability of the photon PDF with respect to the
perturbative order of the QCD calculations used in the fit, in
Fig.~\ref{fig:photoncomp2} we compare the photon PDFs from the
NNPDF3.1luxQED NLO and NNLO fits, normalised to the central value of
the former.
The photon distributions are consistent within uncertainties,
demonstrating good perturbative stability. Indeed, the shift due to
the change in perturbative order is outside the PDF error bands only
in the small-$x$ region, where the photon is sensitive to the
prior PDF used for the computation of the high-$Q^2$ inelastic
component.
In addition, we find that the photon PDF uncertainties are unaffected
by the variation of the perturbative order.

\begin{figure}[t]
\begin{center}
  \includegraphics[scale=0.39]{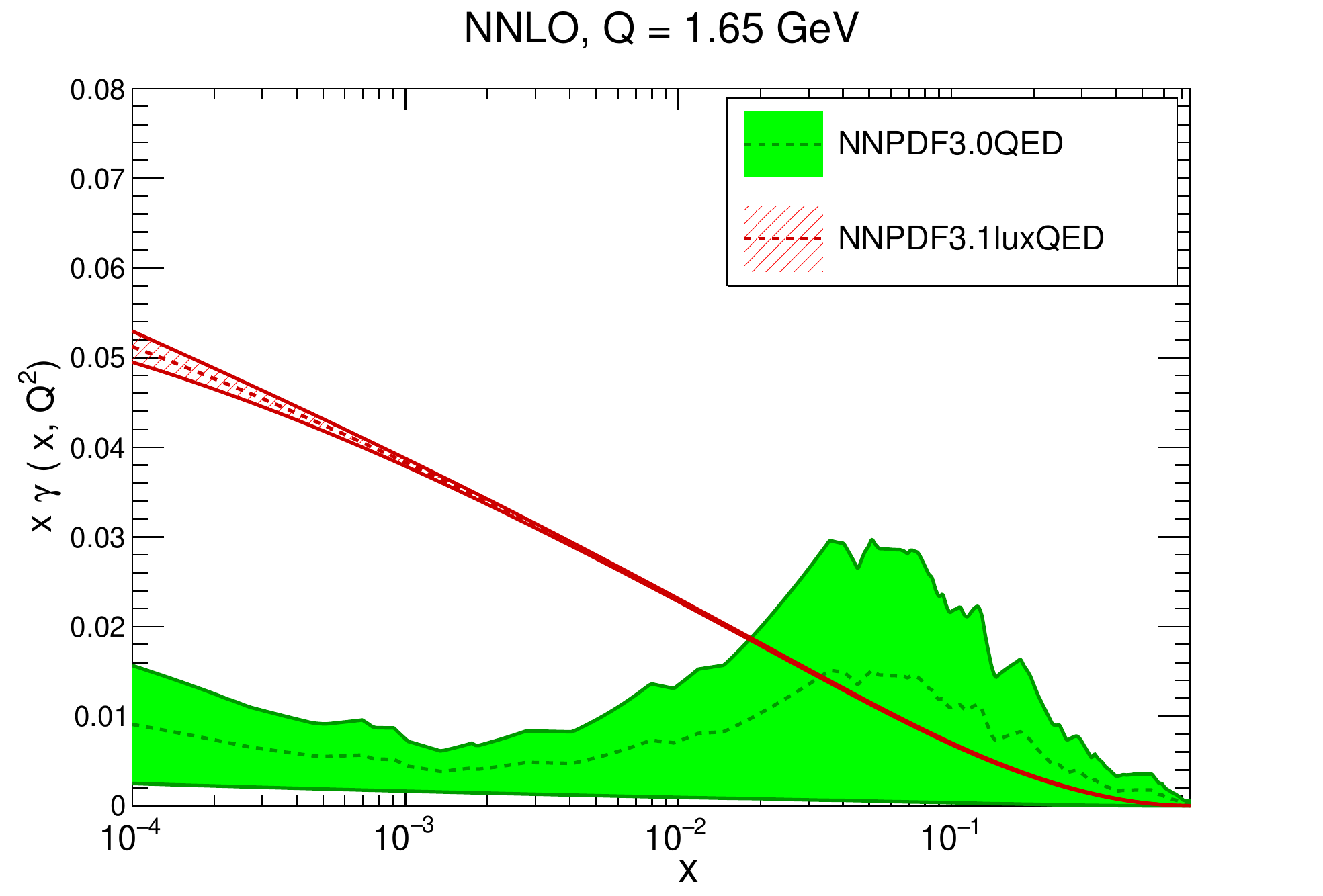}
  \includegraphics[scale=0.39]{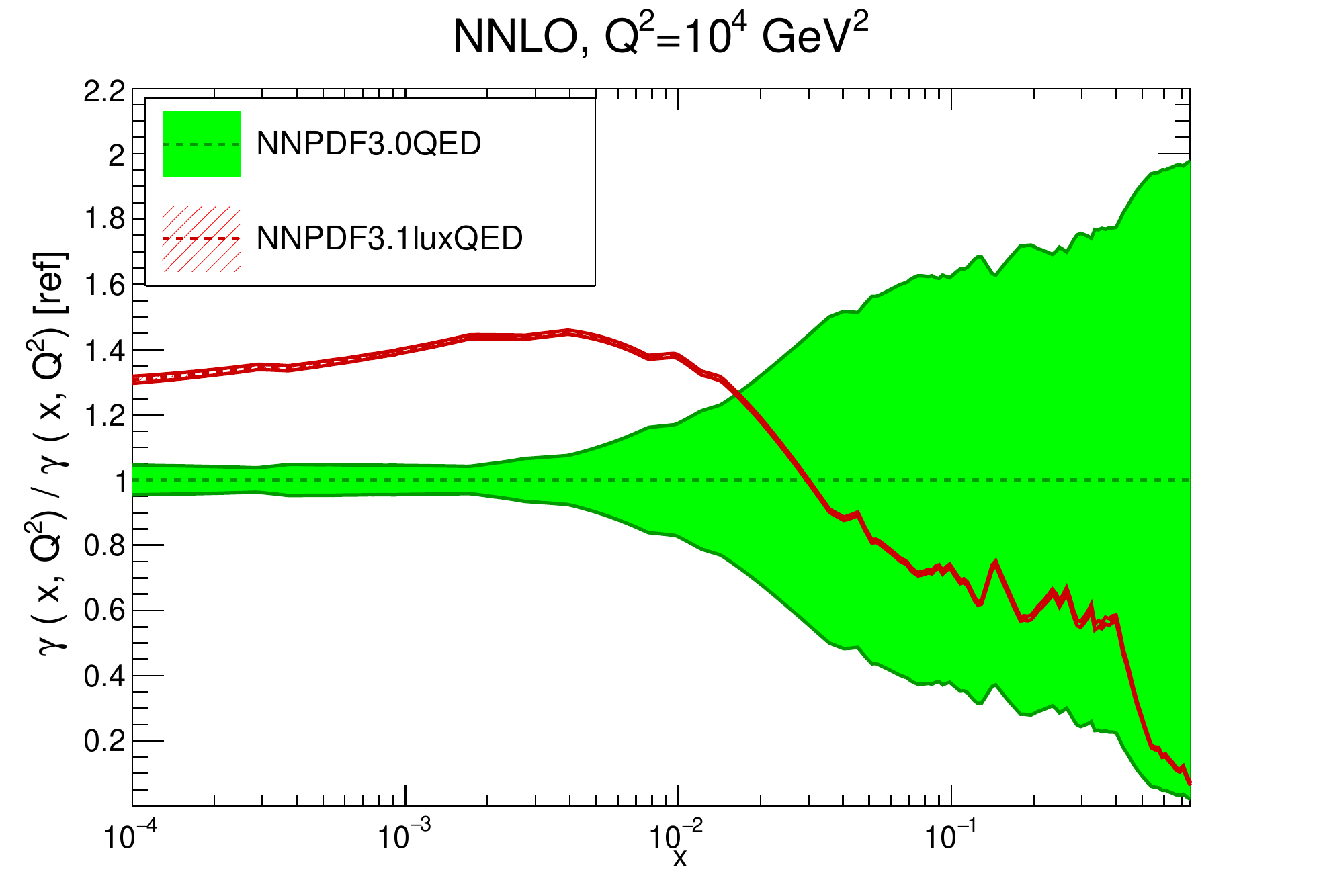}
  \caption{\small Comparison between the photon
    PDF $\gamma(x,Q)$ in NNPDF3.0QED and in NNPDF3.1luxQED
    at $Q=1.65$ GeV (left)
    and at $Q=100$ GeV (right plot).
    In the latter case, results are normalised to
    the central value of NNPDF3.0QED.
     \label{fig:photoncomp30}
   }
\end{center}
\end{figure}

In order to quantify the differences between photon PDFs determined
from global analyses with and without imposing the LUXqed
theoretical constraint, we compare 
NNPDF3.1luxQED with NNPDF3.0QED.
In the following, the PDF uncertainties of NNPDF3.0QED are computed as
68\% confidence-level (CL) intervals, with the central value taken to
be the midpoint of the interval.
In Fig.~\ref{fig:photoncomp30} we show the photon distributions from
these two sets at $Q=1.65$ GeV (left plot) and $Q=100$ GeV (right
plot).
We find that both at low and high scales, in the region
$x\gsim 2\times 10^{-2}$ the two determinations agree within
uncertainties.
For $x\lsim 2\times 10^{-2}$ instead, the NNPDF3.0QED photon
undershoots NNPDF3.1luxQED by up to 40\% at $Q=100$ GeV.
At high scales, PDF uncertainties in NNPDF3.0QED are at the level
of a few percent at small $x$ but become as large as almost 100\% at
large $x$.
The uncertainties in NNPDF3.1luxQED are instead
at the level of a few percent over the entire range in $x$ (see also
Fig.~\ref{fig:photoncomp1}).

As shown in Fig.~\ref{fig:photoncomp30}, for $x\lsim  10^{-2}$ the NNPDF3.0QED photon
undershoots the 3.1luxQED one both at low and at high scales by an amount
which is not covered by the PDF uncertainties of the former.
There are at least two possible contributions to such differences.
First of all, the inclusion of $\mathcal{O}(\alpha^2)$ and $\mathcal{O}(\alpha\alpha_s)$
terms in the DGLAP equations (absent in NNPDF3.0QED), accounts to up to
a difference of 5\% when the photon PDF is evolved from $Q_0=1.65$ GeV to $Q=100$ GeV
(see also Fig.~\ref{fig:lumisQED}), explaining part of the discrepancy.

The second, and more important, potential reason is related to the fact that in
NNPDF2.3QED the boundary condition $\gamma(x,Q_0)$ was determined
from a fit to DIS and Drell-Yan cross-sections
using different settings
for the QCD+QED evolution equations~\cite{Bertone:2016ume} as compared
to those used later to construct NNPDF3.0QED.
This partial mismatch then seems to lead
to a suppression of the photon PDF at small-$x$, explaining
 some of the differences observed in Fig.~\ref{fig:photoncomp30}.
In this context, recall than in NNPDF2.3QED  the
  photon PDF was constrained mostly by the LHC Drell-Yan measurements, which makes
  tricky the mapping between how different evolution settings translate into a change in the fitted
boundary condition.
In any case, is clear that pinning down the underlying origin of these differences for
  $x\lsim  10^{-2}$ would require redoing the NNPDF2.3QED fit with exactly
  the same theoretical settings for the DGLAP evolution
  as in NNPDF3.1QED, which is beyond the scope of this paper.

\begin{figure}[t]
\begin{center}
\includegraphics[scale=0.39]{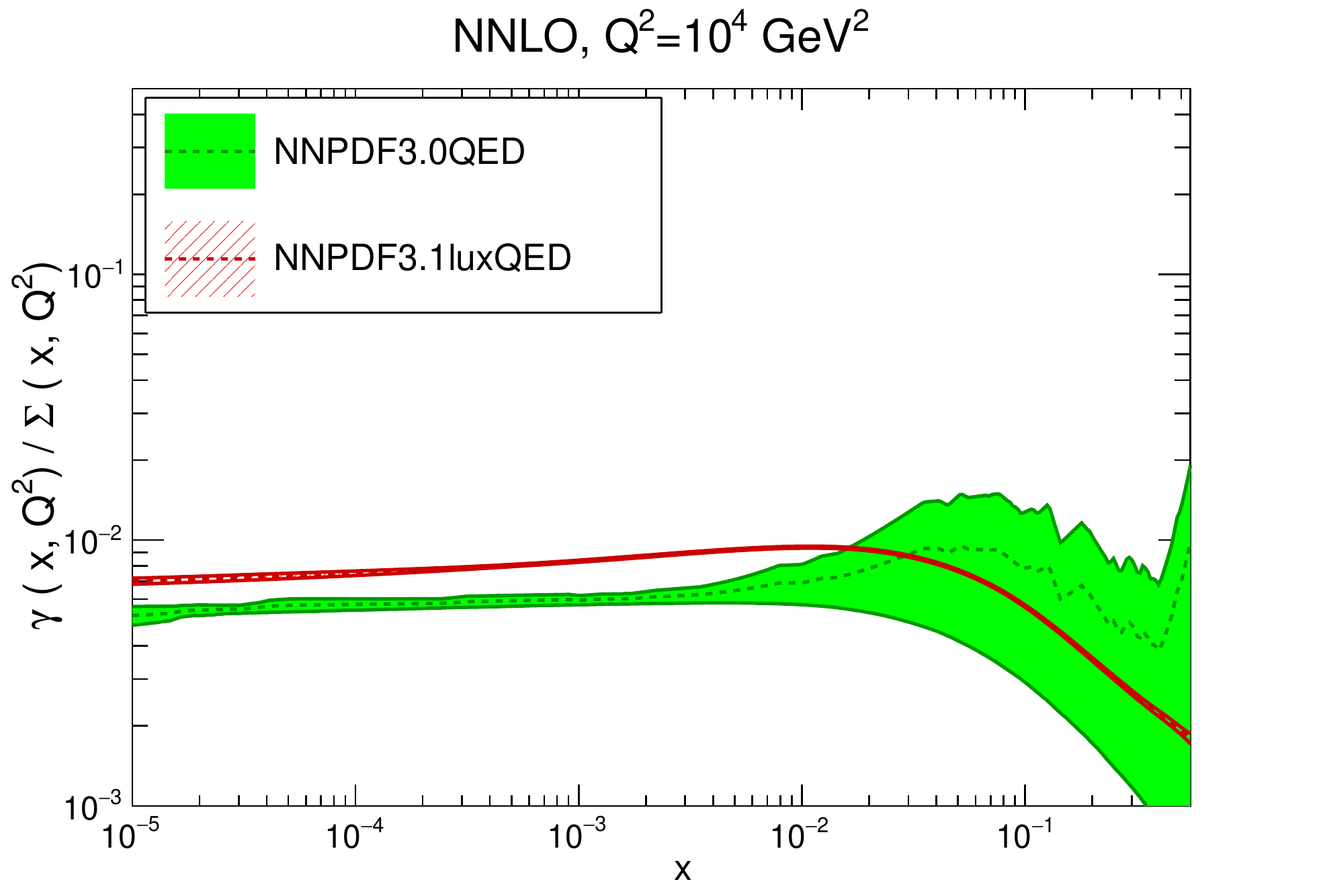}
\includegraphics[scale=0.39]{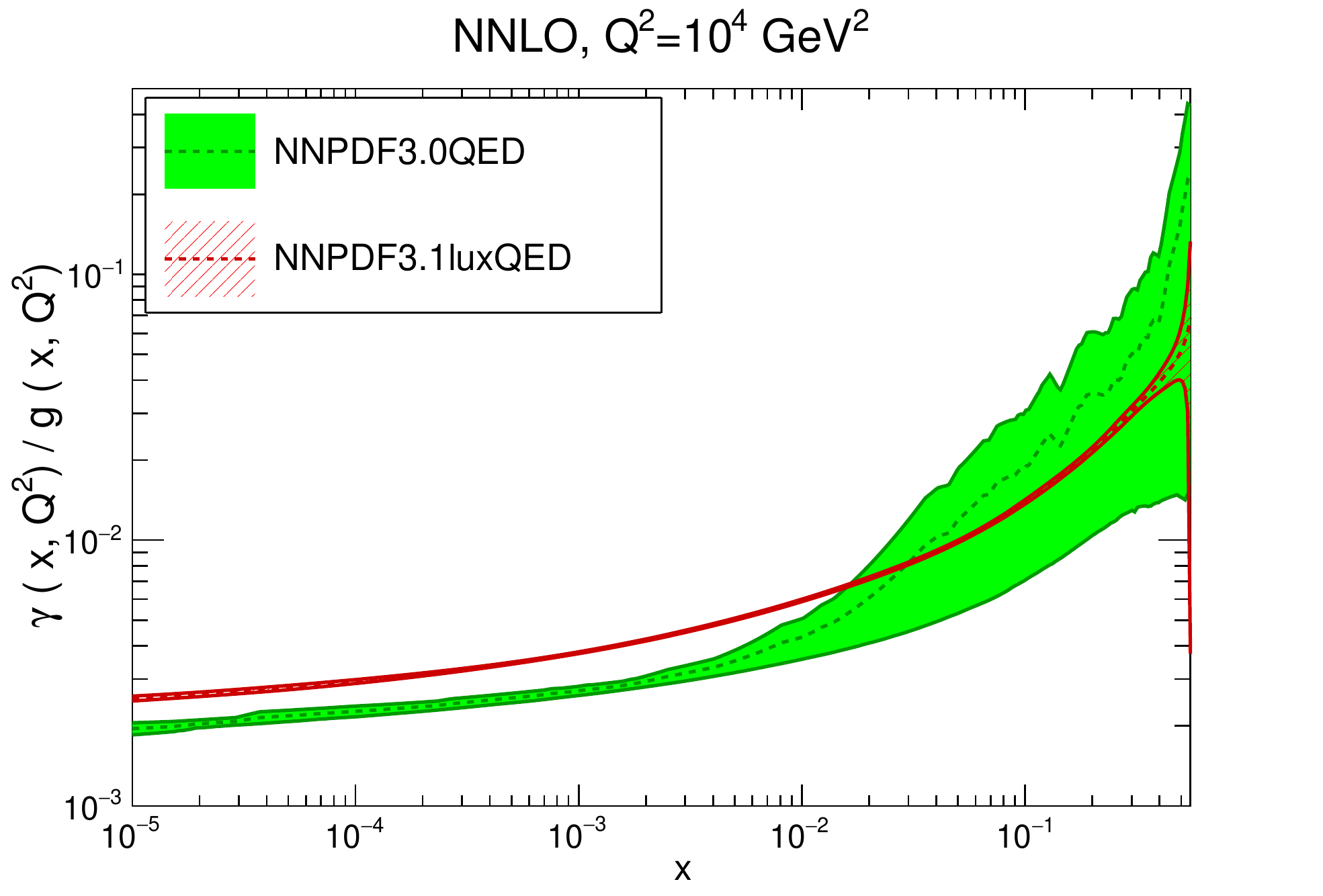}
   \caption{\small The ratios of the photon PDF to the quark singlet
     $\gamma(x,Q)/\Sigma(x,Q)$ (left) and to the gluon
     $\gamma(x,Q)/g(x,Q)$ PDFs (right)
     for $Q=100$ GeV, comparing NNPDF3.0QED and NNPDF3.1luxQED.
     The corresponding LUXqed17 results are very similar
     to the NNPDF3.1luxQED ones and thus not shown.
     \label{fig:photonratios}
  }
\end{center}
\end{figure}

It is also interesting to examine the relative size of the photon PDF
with respect to the total quark singlet and gluon PDFs. This allows us
to estimate where PI contribution to hadron-collider processes becomes
sizeable as compared to quark- and gluon-initiated subprocesses.
In Fig.~\ref{fig:photonratios} we compare the predictions of
NNPDF3.0QED and NNPDF3.1luxQED for the $\gamma(x,Q)/\Sigma(x,Q)$
(left) and $\gamma(x,Q)/g(x,Q)$ (right) ratios at $Q=100$ GeV.
From the right panel of Fig.~\ref{fig:photonratios} we observe that
the $\gamma/\Sigma$ ratio is around
$\mathcal{O}(10^{-2})\simeq \mathcal{O}(\alpha_{\rm QED})$ over the
entire range of $x$.
On the other hand, from the right panel of Fig.~\ref{fig:photonratios}
we find that for $x\gsim 0.01$ the $\gamma/g$ ratio becomes larger
than the $\gamma/\Sigma$ one. In fact, the $\gamma/g$ ratio is as
large as 10\% for $x \simeq 0.5$.

Therefore, we find that that complementing a data-driven determination
of the photon PDF with the LUXqed theoretical constraints  allows for a precise
determination of the photon PDF in most of the range of $x$ relevant
for applications at the LHC.

\subsection{QED effects on the quark and gluon PDFs}
\label{sec:quarkgluoncomparison}

In this section we study the quark and gluon PDFs in NNPDF3.1luxQED as
compared to their corresponding QCD-only counterparts in NNPDF3.1.
This comparison gauges the impact on quarks and gluons of three
different QED effects: the modification of the momentum sum rule, the
QED splitting functions in the DGLAP evolution equations, and the QED
corrections to the DIS coefficient functions.

\begin{figure}[t]
  \begin{center}
\includegraphics[scale=0.39]{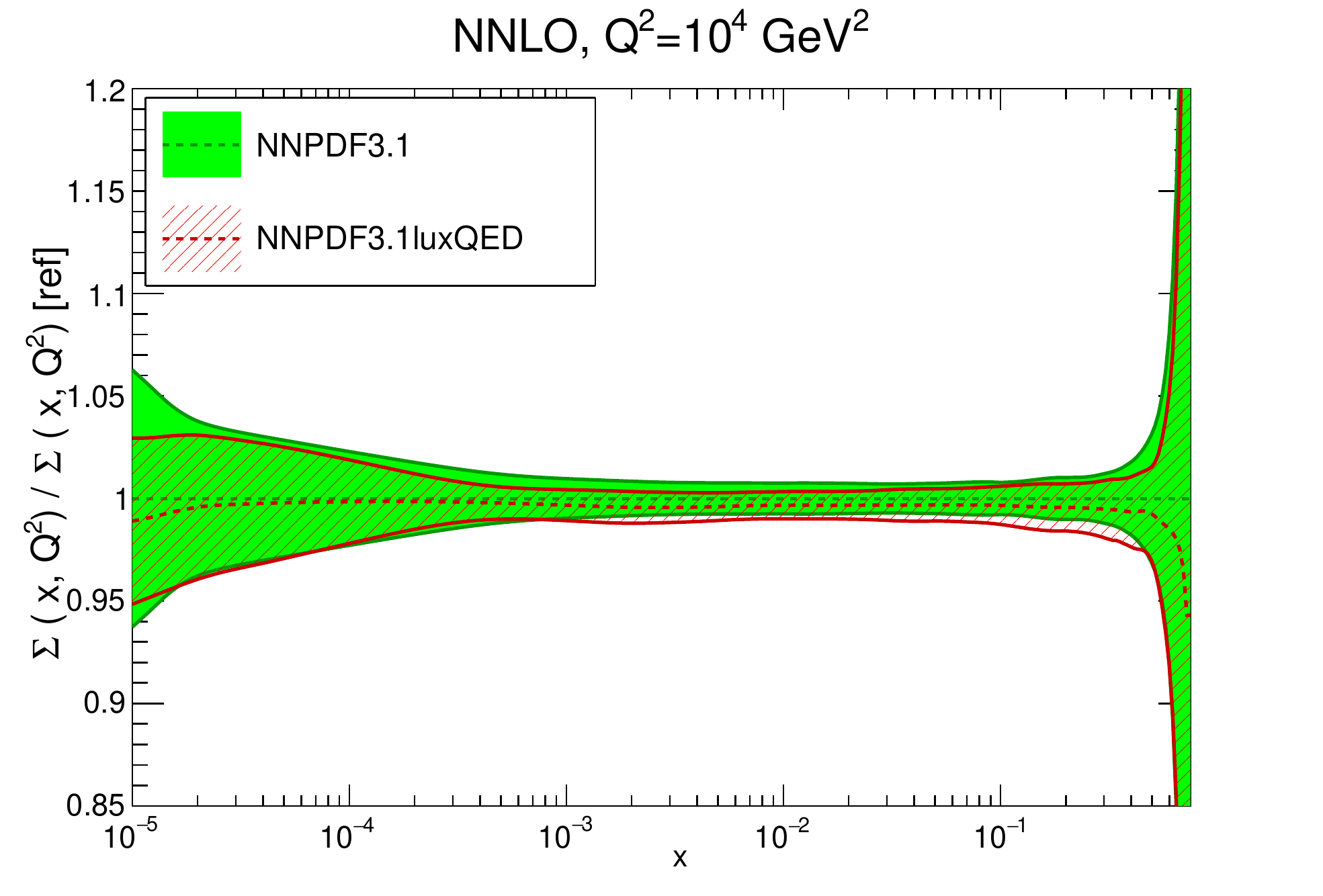}
\includegraphics[scale=0.39]{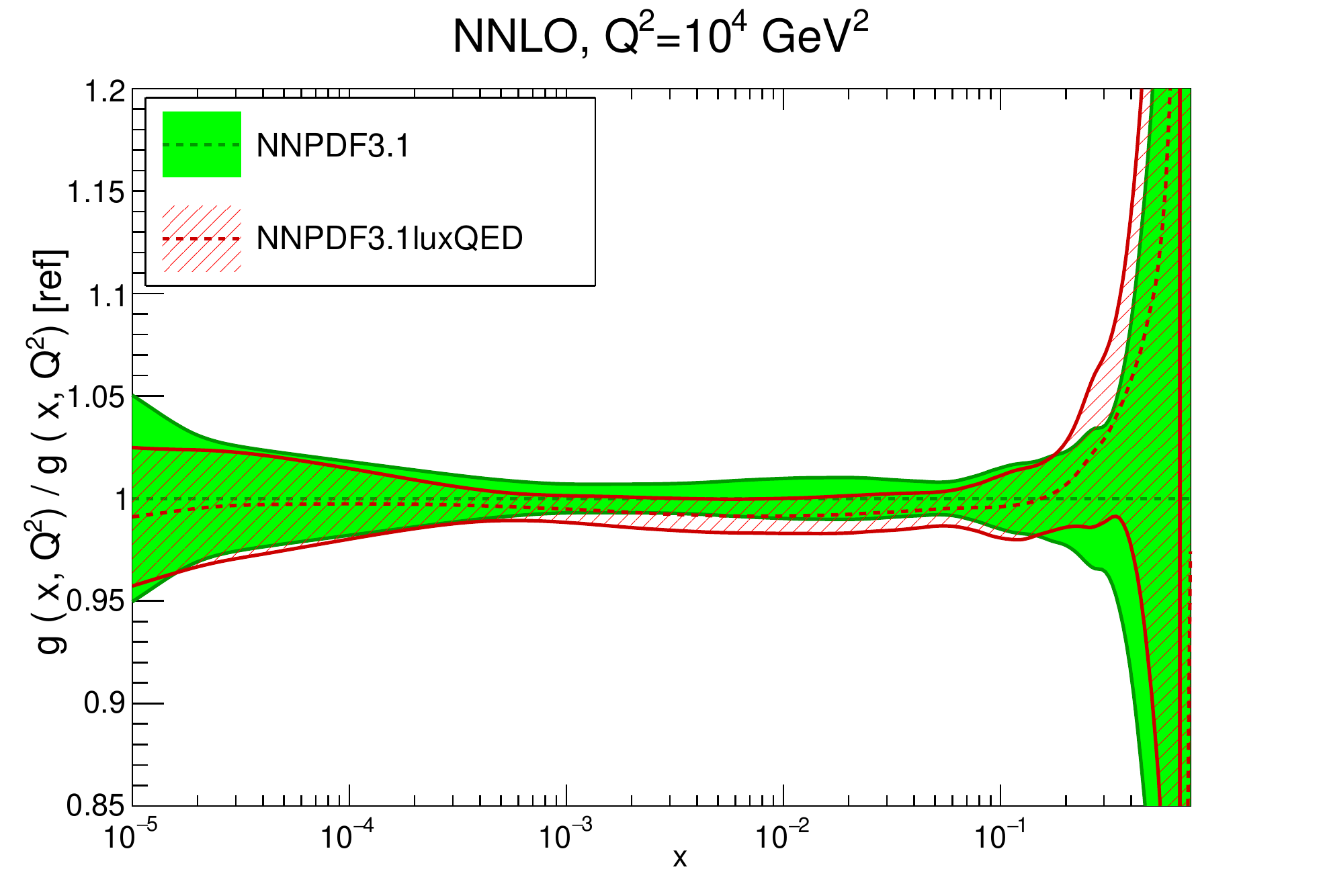}
  \caption{\small Comparison of the total quark singlet (left)
    and gluon PDFs (right) between NNPDF3.1 and NNPDF3.1luxQED at
    $Q=100$ GeV, normalised to the central value of
    the former.
    \label{fig:qcd_nnpdf31qed-vs-nnpdf31}
  }
\end{center}
\end{figure}

In Fig.~\ref{fig:qcd_nnpdf31qed-vs-nnpdf31} we show the singlet and
gluon PDFs of the NNPDF3.1 and NNPDF3.1luxQED sets at $Q=100$ GeV
normalised to the central value of the former.
While differences at the level of the singlet are small, differences for the
gluon PDF are somewhat larger.
Indeed, the NNPDF3.1luxQED gluon is smaller than its QCD counterpart
by about 1\% at $x\simeq 10^{-2}$ and enhanced by about $5\%$ for
$x\simeq 0.5$.
In both cases, the shift in the central values is at the edge of the
corresponding PDF uncertainty band.
The effect on the gluon PDF can be explained by observing that, as we
will discuss in Sect.~\ref{sec:msr}, the photon PDF can carry up to
0.5\% of the proton momentum. This fraction is effectively subtracted
from the singlet and gluon distributions by means of the sum rule,
Eq.~(\ref{sec:momentumsumrule}).
However, the sum rule mostly affects the gluon PDF because the
normalisation of the quark singlet is more tightly constrained from the
DIS inclusive structure function data.
We conclude that the back-reaction of QED effects onto the quark and
gluon PDFs is small but not negligible, particularly for the latter.

\begin{figure}[t]
  \begin{center}
\includegraphics[scale=0.39]{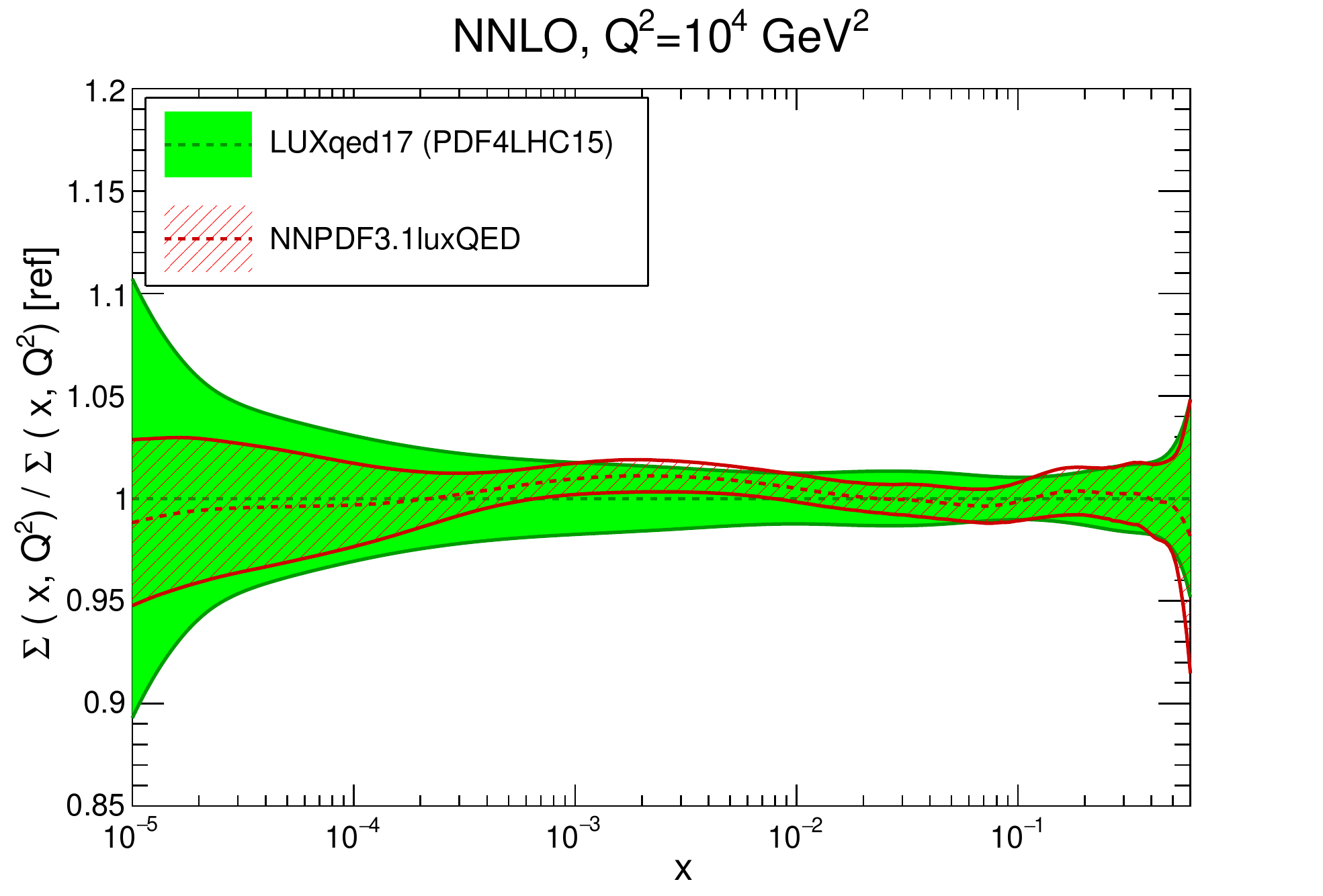}
\includegraphics[scale=0.39]{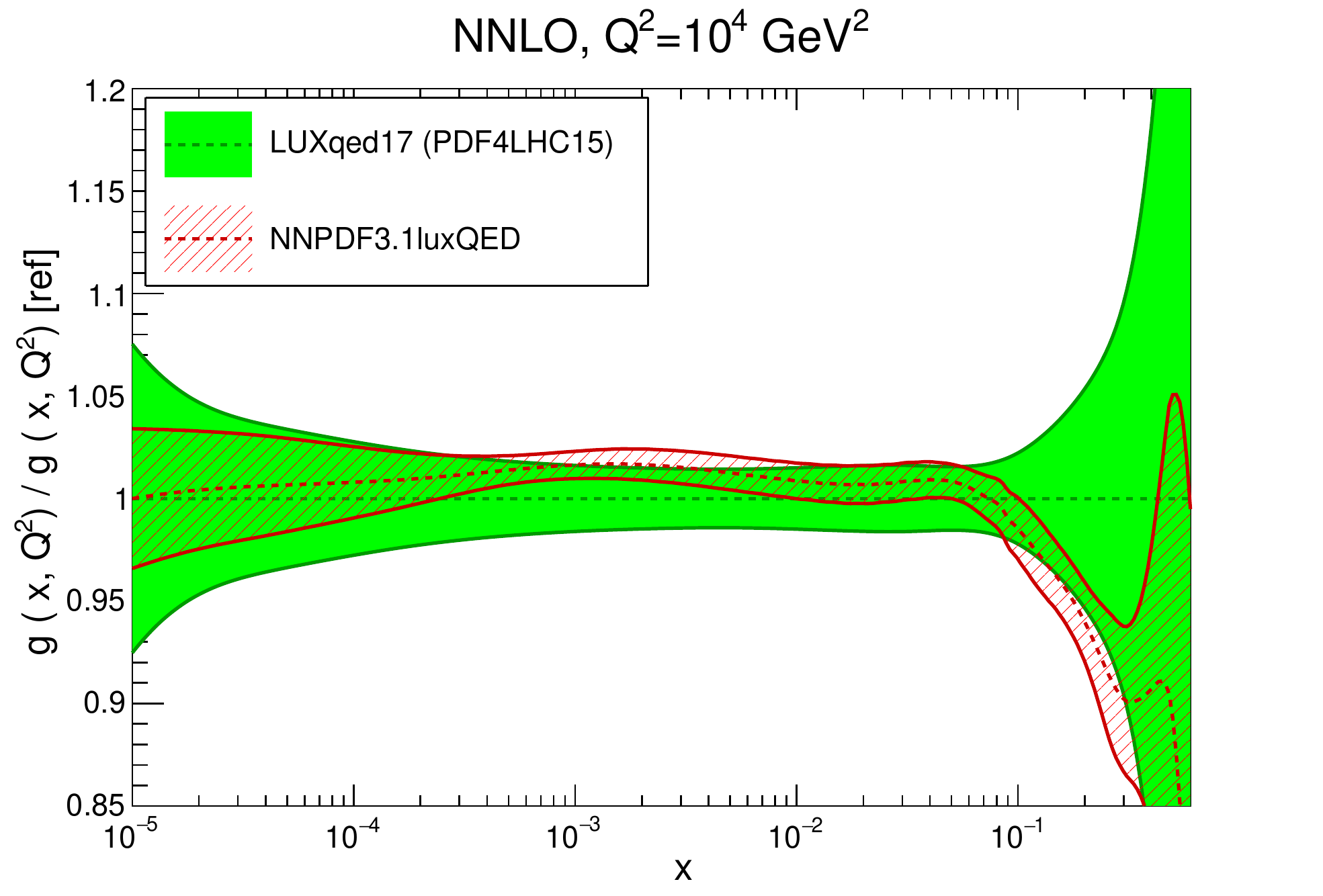}
\caption{\small Same as Fig.~\ref{fig:qcd_nnpdf31qed-vs-nnpdf31},
  now comparing the quark singlet and gluon of NNPDF3.1luxQED with those
  of LUXqed17, which correspond closely to the PDF4LHC15 NNLO set.
    \label{fig:qcd_nnpdf31qed-vs-luxqed17}
  }
\end{center}
\end{figure}

For completeness, in Fig.~\ref{fig:qcd_nnpdf31qed-vs-luxqed17} we show
the same comparison as in Fig.~\ref{fig:qcd_nnpdf31qed-vs-nnpdf31} but
now between NNPDF3.1luxQED and LUXqed17.
Note that the quark and gluon PDFs
of LUXqed17 correspond closely to those of the PDF4LHC15 set,
differing only by a rescaling of the gluon PDF and by the QED contributions
to the DGLAP evolution.
Since the PDF4LHC15 set is built as a combination of three
different PDF sets, namely CT14, MMHT14, and NNPDF3.0, it
exhibits larger uncertainties that the individual sets.
We find good compatibility between the singlet and gluon of NNPDF3.1luxQED
and LUXqed17, with the PDF errors of the former being rather smaller.
These reduced uncertainties are particularly noticeable for the medium and
large-$x$ gluon PDF, due to the several gluon-sensitive experiments included in
NNPDF3.1 such as top-quark pair distributions~\cite{Czakon:2016olj} and
the $Z$ boson $p_T$~\cite{Boughezal:2017nla}.

\subsection{Partonic luminosities}

\begin{figure}[t]
\begin{center}
  \includegraphics[scale=0.38]{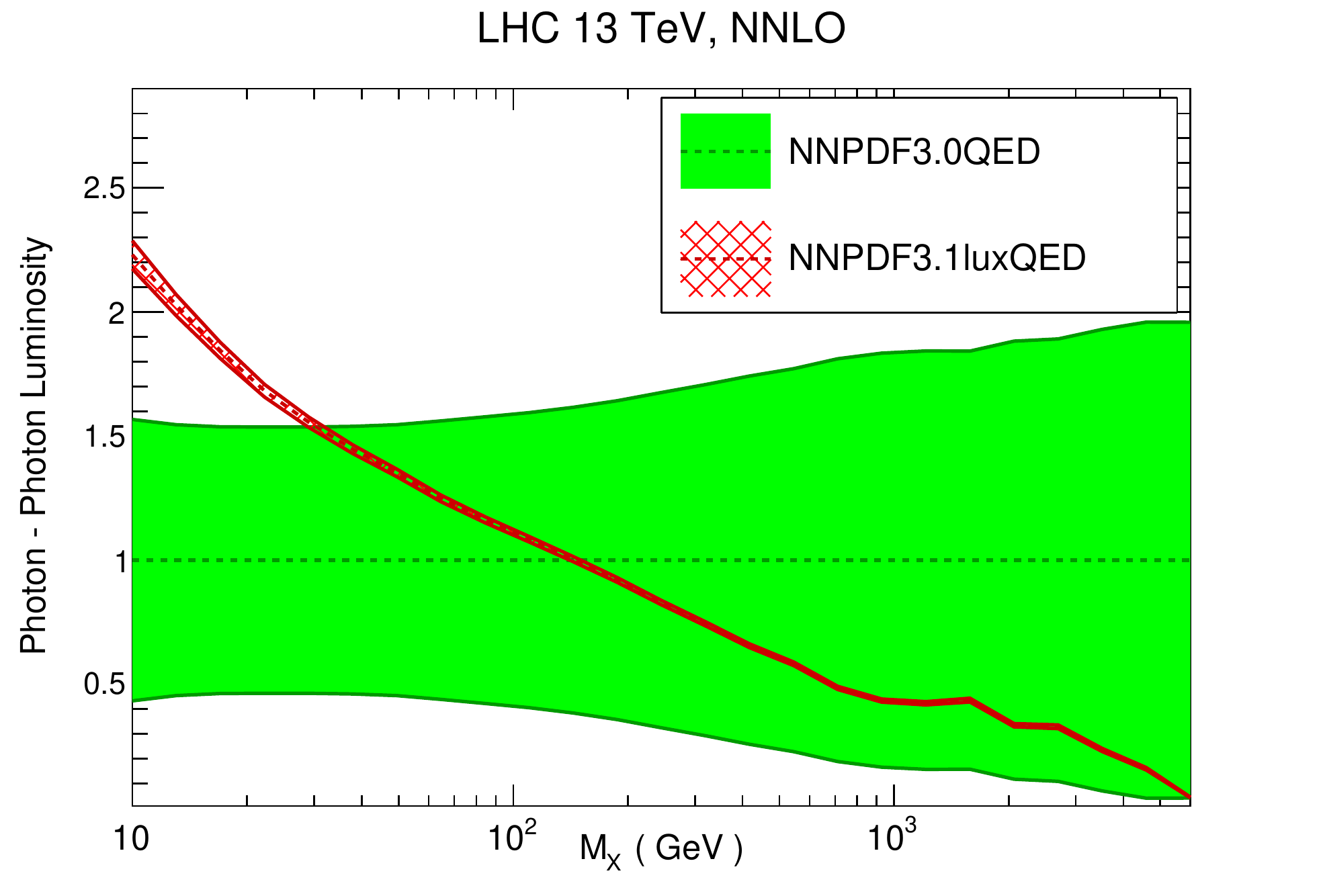}
  \includegraphics[scale=0.38]{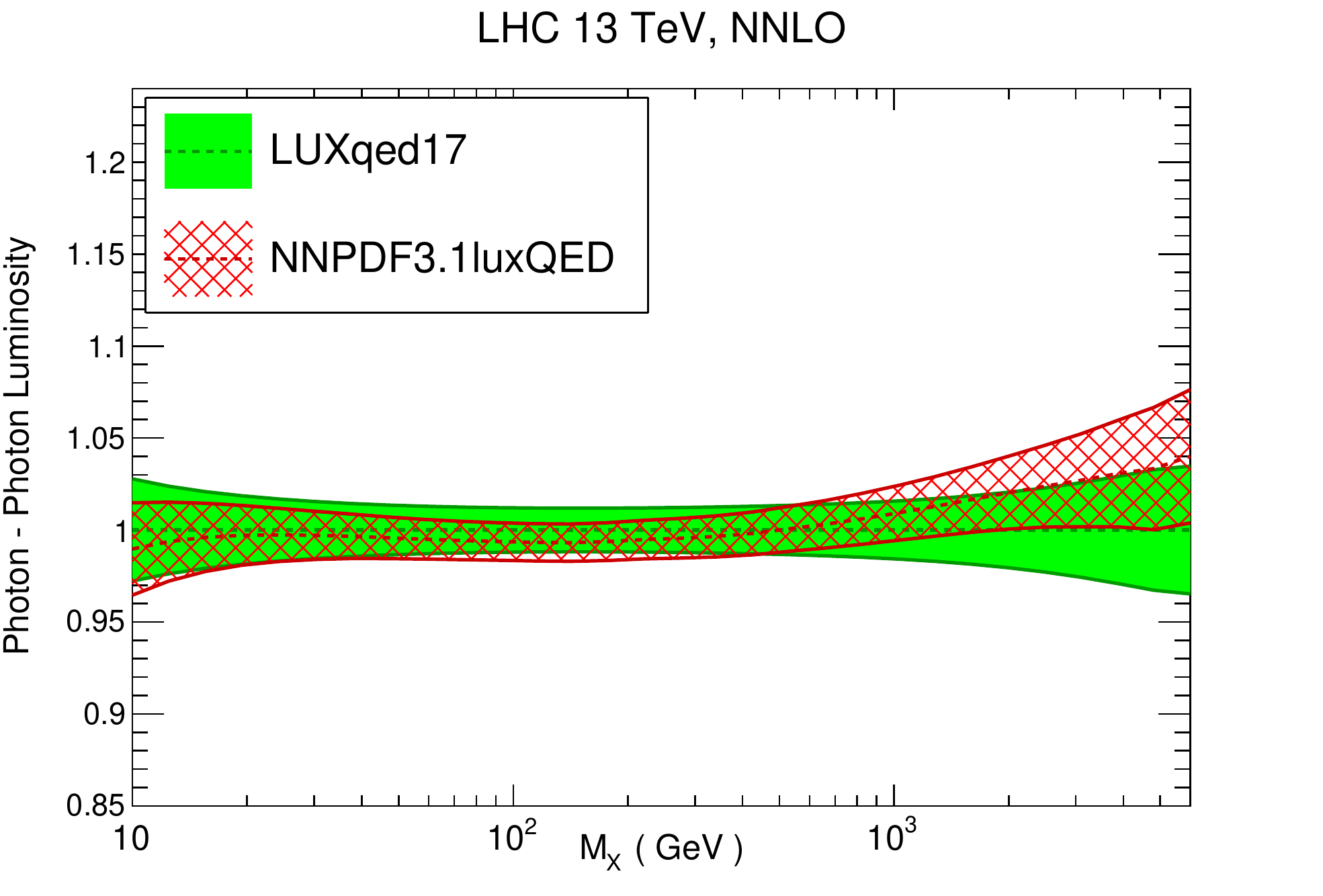}
  \caption{\small Comparison of the $\gamma\gamma$
    PDF luminosities between NNPDF3.1luxQED
    and NNPDF3.0QED (left) and LUXqed17 (right plot) as a
    function of $M_X$ for $\sqrt{s}=13$ TeV.
    Note that the $y$ axis range is different in the both
    plots.
    \label{fig:phenoLumi1}
  }
\end{center}
\end{figure}

Next we compare partonic luminosities integrated over
rapidity as a function of the final-state invariant mass
$M_X$ for photon-photon and photon-quark
initial states (see~\cite{Mangano:2016jyj} for the definitions used).
In Fig.~\ref{fig:phenoLumi1} we compare the
$\mathcal{L}_{\gamma\gamma}$ luminosity obtained with the
NNPDF3.0QED, NNPDF3.1luxQED, and LUXqed17 sets for a centre-of-mass
energy of $\sqrt{s}=13$ TeV.
From the left panel of Fig.~\ref{fig:phenoLumi1} we observe good agreement
between NNPDF3.0QED and NNPDF3.1luxQED.
The two sets agree within uncertainties over the entire mass range considered,
except for $\mathcal{L}_{\gamma\gamma}$ at $M_X\lsim 30$ GeV where
NNPDF3.1luxQED overshoots NNPDF3.0QED.

Evident once again is the effect of the LUXqed constraints on the
photon PDF uncertainty, with the errors of just a few percent as
compared to the determination that does not account for them.
From Fig.~\ref{fig:phenoLumi1} we also see that there is good
agreement between LUXqed17 and NNPDF3.1luxQED both at the level of central
values and  of uncertainties.
Only at $M_X\gsim 1$ TeV NNPDF3.1luxQED
tends to be a few percent larger than LUXqed17.
Similar considerations hold for the $\mathcal{L}_{q\gamma}$ luminosities.

\begin{figure}[t]
\begin{center}
  \includegraphics[scale=0.38]{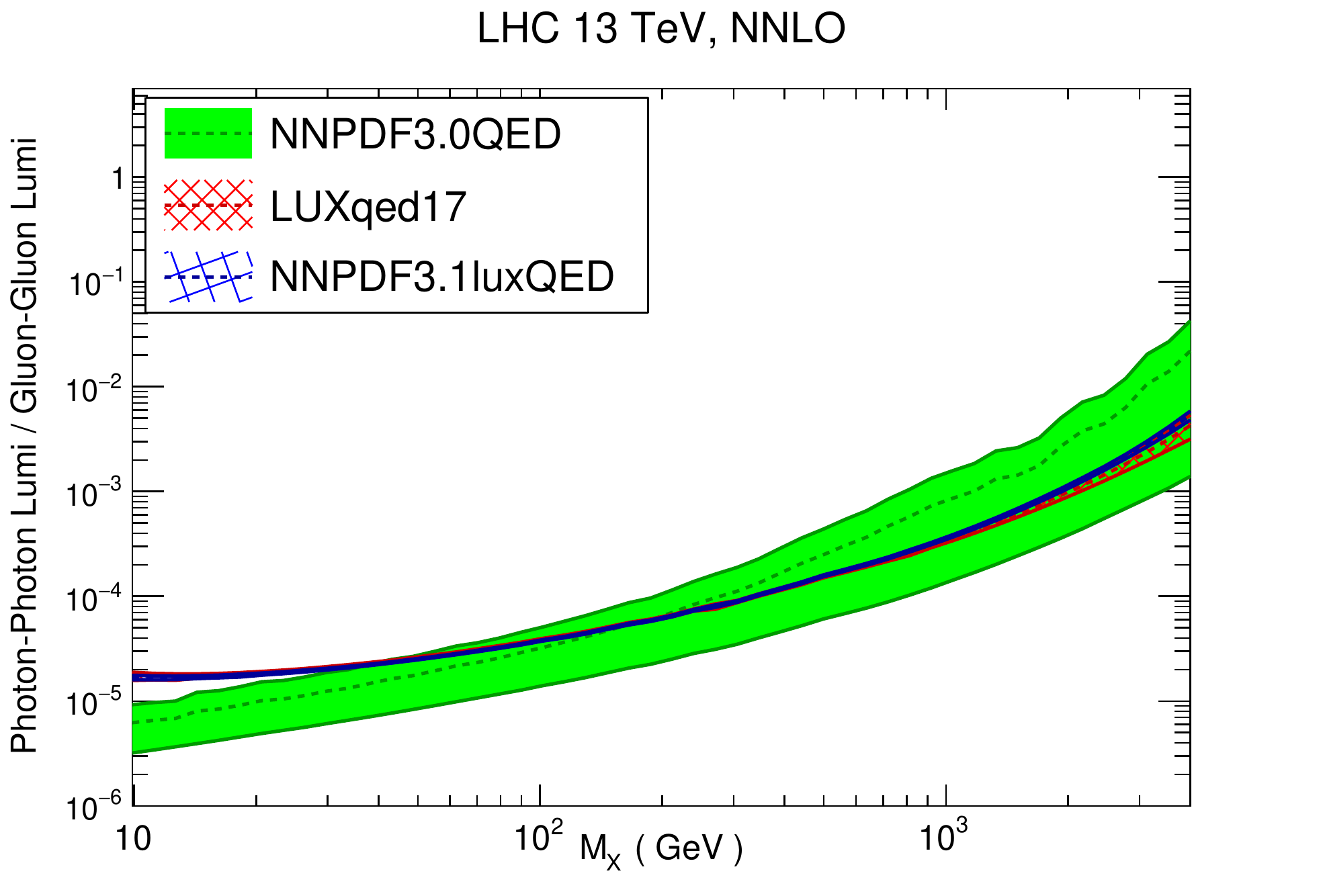}
  \includegraphics[scale=0.38]{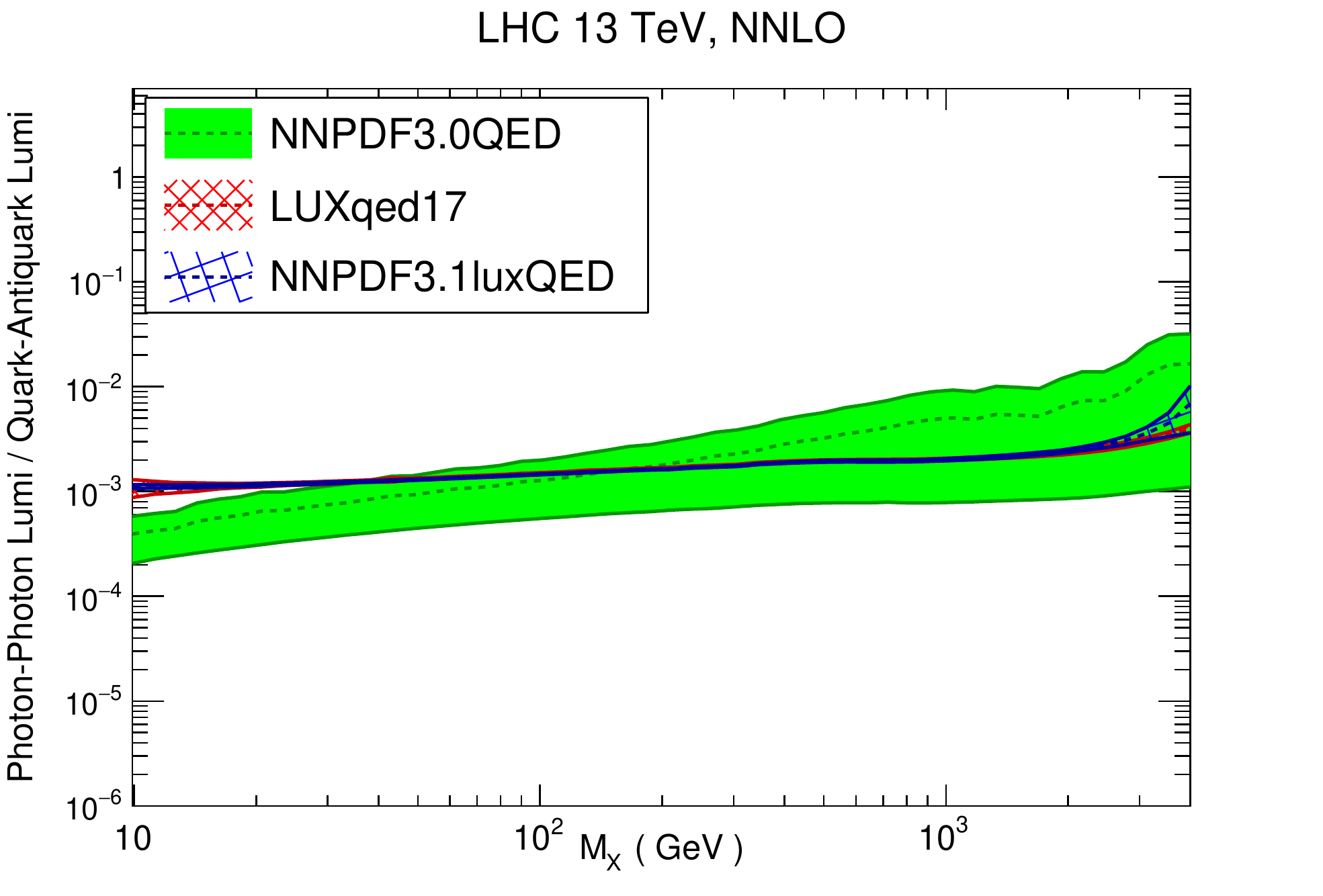}
  \caption{\small The $\mathcal{L}_{\gamma\gamma}/\mathcal{L}_{gg}$ (left)
    and $\mathcal{L}_{\gamma\gamma}/\mathcal{L}_{q\bar{q}}$ (right plot) ratios
    of PDF luminosities.
    \label{fig:phenoLumiRat}
  }
\end{center}
\end{figure}

Following the PDF-level comparisons presented in
Sect.~\ref{sec:PDFcomparisonPhoton}, it is instructive to also consider the
ratios of the photon-photon luminosity over gluon-gluon and over
quark-antiquark luminosities,
$\mathcal{L}_{\gamma\gamma}/\mathcal{L}_{gg}$ and
$\mathcal{L}_{\gamma\gamma}/\mathcal{L}_{q\bar{q}}$.
These ratios are interesting since they provide an estimate of the
relative importance of the PI contribution over quark- and
gluon-initiated contributions as a function of $M_X$.
These ratios are shown in Fig.~\ref{fig:phenoLumiRat}, where we
compare the results from NNPDF3.0QED, NNPDF3.1luxQED, and LUXqed17.
The $\mathcal{L}_{\gamma\gamma}/\mathcal{L}_{q\bar{q}}$ ratio
varies very mildly with $M_X$, with a value $\simeq 10^{-3}$.
On the other hand, the ratio $\mathcal{L}_{\gamma\gamma}/\mathcal{L}_{gg}$
increases steeply with the final state invariant mass $M_X$, beginning at $\simeq
10^{-5}$ at low invariant masses and growing up to $\simeq 10^{-3}$ for $M_X=4$
TeV.
For large invariant masses, we therefore find that the two ratios
$\mathcal{L}_{\gamma\gamma}/\mathcal{L}_{gg}$ and
$\mathcal{L}_{\gamma\gamma}/\mathcal{L}_{q\bar{q}}$ take similar
values.
For both cases, the results of the three PDF sets are consistent
within uncertainties, with NNPDF3.1luxQED and LUXqed17 exhibiting
significantly reduced errors as compared to NNPDF3.0QED.

\subsection{The momentum fraction carried by photons}\label{sec:msr}

\begin{figure}[t]
\begin{center}
  \includegraphics[scale=0.39]{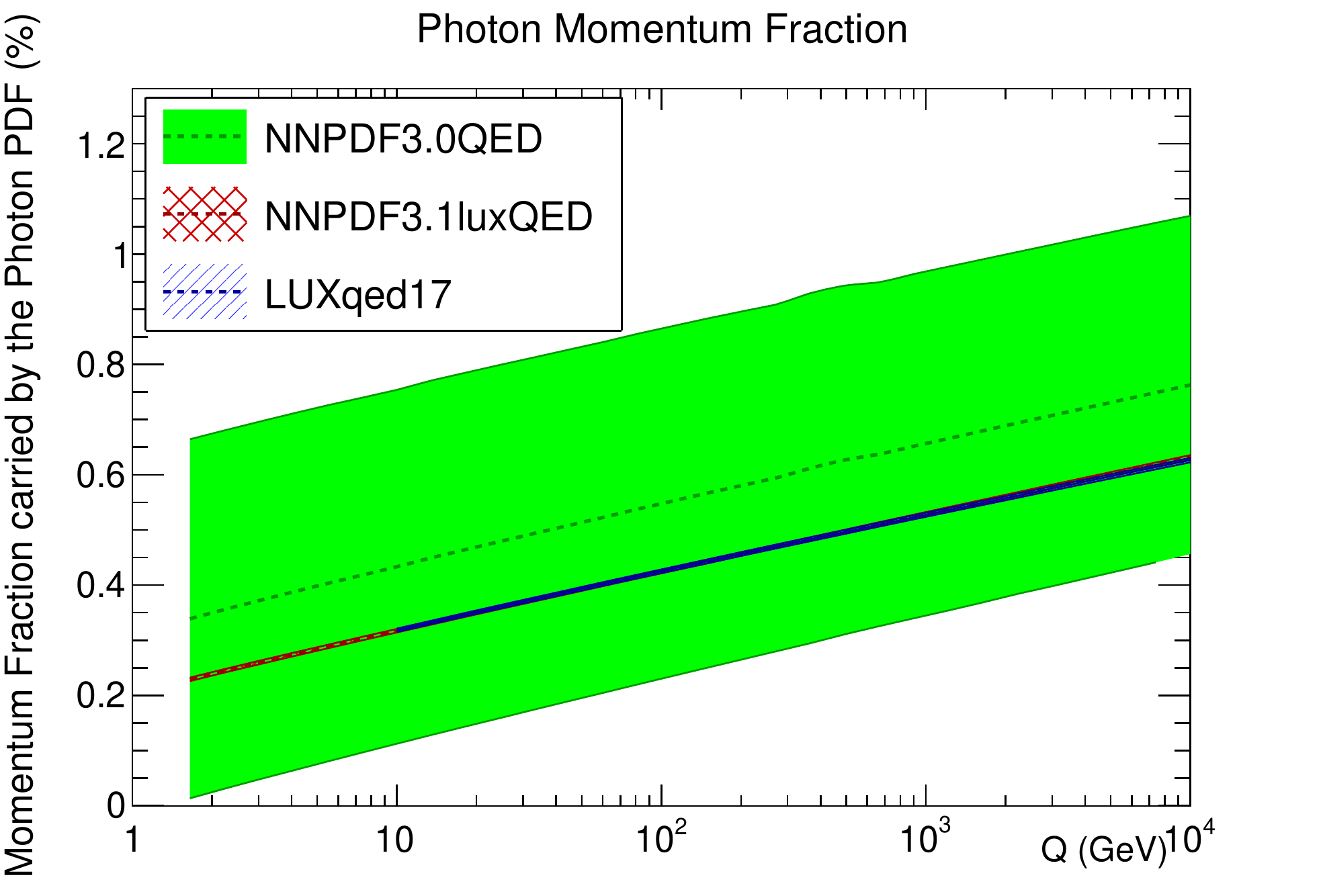}
  \includegraphics[scale=0.39]{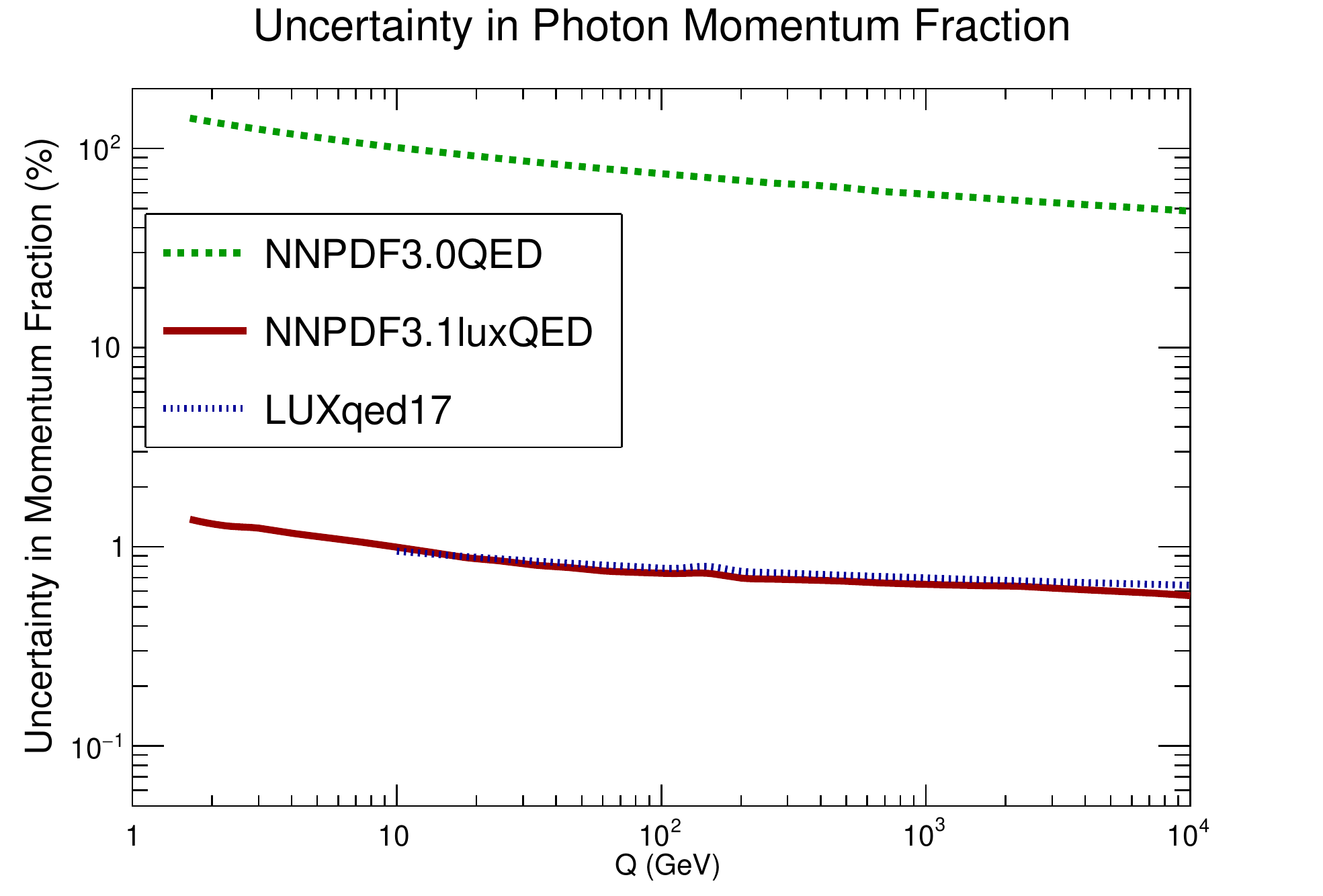}
  \caption{\small The momentum fraction $\la x\ra_\gamma$
    carried
    by photons in the proton (left)
    and its percentage uncertainty (right) as a function of 
    $Q$ for NNPDF3.0QED, NNPDF3.1luxQED, and
LUXqed17.
    \label{fig:photonmomsunmrule} } 
\end{center} 
\end{figure}

As the photon PDF carries a non-zero amount of the total proton
momentum, it therefore contributes to the momentum sum rule,
Eq.~(\ref{sec:momentumsumrule}).
Here we examine the momentum fraction carried by the photon PDF
\begin{equation}
  \label{eq:photonmomfrac}
    \la x\ra_\gamma(Q) \equiv
\int_0^1\,dx\,x\gamma(x,Q)\,. 
\end{equation} 
In Fig.~\ref{fig:photonmomsunmrule} we show the value of
$\la x\ra_\gamma$ in the NNPDF3.1luxQED, NNPDF3.0QED, and LUXqed17
sets as a function of the scale $Q$. In the right plot of
Fig.~\ref{fig:photonmomsunmrule} we also show the corresponding
percent PDF uncertainties.
For LUXqed17 we restrict the comparison to the validity region of this
set, \textit{i.e.} $Q \ge 10$ GeV.

From Fig.~\ref{fig:photonmomsunmrule} we observe that, while the
NNPDF3.0QED determination is affected by large PDF uncertainties, the
other two sets lead to a compatible prediction for the photon momentum
fraction for all values of $Q$.
As expected from the PDF-level comparisons, there is a significant
reduction in the uncertainty on the value of $\la x\ra_\gamma$ once
the LUXqed theoretical constraints are accounted for.
Indeed, while in NNPDF3.0QED the uncertainties in the photon momentum
fraction range from around 50\% to 100\%, in NNPDF3.1luxQED the
contribution of the photon PDF to the momentum of the proton is known
with an accuracy better than 1\% over the entire range in $Q$.
Nevertheless, the central value of $\la x\ra_\gamma$ in NNPDF3.0QED
turns out to be rather close to that of NNPDF3.1luxQED, highlighting
the consistency between the two approaches.

\begin{table}[h!]
  \centering
   \renewcommand{\arraystretch}{1.25}
  \begin{tabular}{c|c|c}
   \toprule
    &        $\la x\ra_\gamma(Q=1.65\,{\rm GeV})$   &
    $\la x\ra_\gamma(Q=m_Z)$ \\
    \midrule
  NNPDF3.0QED    &   $\lp 0.3 \pm 0.3\rp $\%  &   $\lp 0.5 \pm 0.3\rp$\% \\

  NNPDF3.1luxQED  &  $\lp 0.229 \pm 0.003\rp$\%  &  $\lp 0.420 \pm 0.003\rp$\% \\

  LUXqed17    & $-$   &     $\lp 0.421 \pm 0.003\rp\% $ \\
  \bottomrule
  \end{tabular}
  \vspace{0.3cm}
  \caption{\small
    \label{tab:momentumfraction}
    The momentum fraction $\la x\ra_\gamma$
    carried
    by photons in the proton, 
Eq.~(\ref{eq:photonmomfrac}),
at the initial parametrization scale $Q=Q_0=1.65$ GeV and at typical LHC scale $Q=m_Z$.}
\end{table}

In Tab.~\ref{tab:momentumfraction} we report the photon momentum
fraction Eq.~(\ref{eq:photonmomfrac}) both at the initial parametrisation
scale $Q_0=1.65$ GeV and at $Q=m_Z$ for the three PDF sets including
the associated uncertainties.
While in NNPDF3.0QED the photon momentum fraction at the initial scale
is consistent with zero, in NNPDF3.1luxQED one finds a non-zero photon
momentum fraction with very high statistical significance.
In particular, the photon momentum fraction in NNPDF3.1luxQED
increases from $0.23\%$ at low scales to $0.42\%$ at $Q=m_Z$, with
small uncertainties in both cases.
For $Q=m_Z$, the results of NNPDF3.1luxQED are fully consistent with
those of LUXqed17, as also shown in Fig.~\ref{fig:photonmomsunmrule}.

\section{Photon-initiated processes at the LHC}
\label{sec:pheno}

\begin{figure}[t]
\begin{center}
  \includegraphics[width=\textwidth]{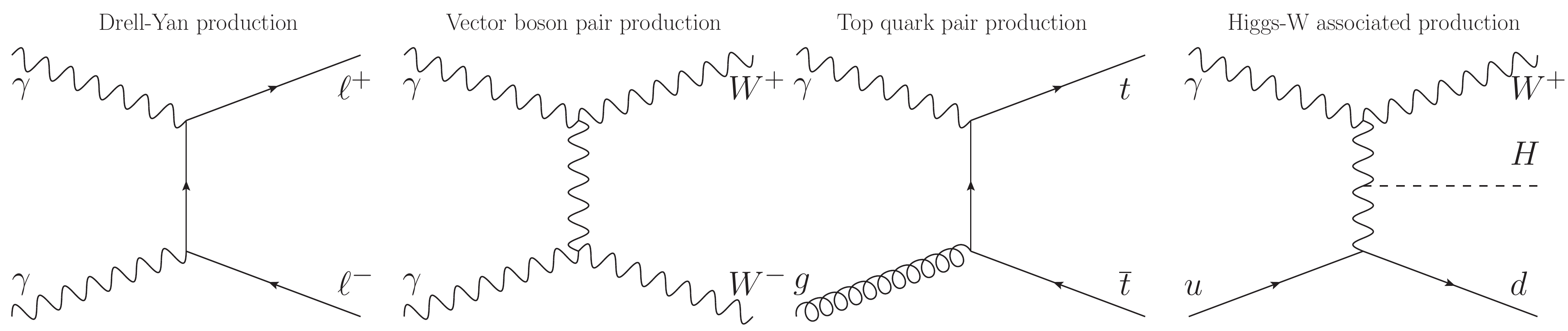}
  \caption{\small Representative PI diagrams for various LHC processes: 
    Drell-Yan, vector-boson pair production, top-quark pair
    production, and the associated production of a Higgs with a $W$ boson.
    \label{fig:photoninitiated}
  }
\end{center}
\end{figure}

We shall now explore some of the implications of NNPDF3.1luxQED for LHC
phenomenology.
Specifically, we shall investigate the application of this new
set to the study of Drell-Yan, vector-boson pair production, top-quark pair
production, and the associated production of a Higgs boson with a $W$ boson.
Representative PI diagrams contributing to these processes at the Born level are shown in
Fig.~\ref{fig:photoninitiated}.
Our aim is to assess the relative size of the PI contributions with respect to
quark- and gluon-initiated subprocesses at the $\sqrt{s}=13$ TeV LHC.
See also Refs.~\cite{Bertone:2015lqa,Schmidt:2015zda,Mangano:2016jyj,
  Harland-Lang:2016qjy,Bourilkov:2016qum,Accomando:2016tah,Pagani:2016caq}
for recent studies.

The results presented in this section have been obtained at leading order in both
the QCD and QED couplings using {\tt MadGraph5\_aMC@NLO} interfaced to
{\tt APPLgrid} through {\tt aMCfast}.
We have used the default values
in {\tt MadGraph5\_aMC@NLO} v2.6.0 for the couplings and other
electroweak parameters, as defined in the 
standard model setup.
In particular, we use the default value $\alpha=1/132.51$ for the QED
coupling and ignore the effects of its running that are beyond the
accuracy of the calculation.

We will compare the predictions of NNPDF3.1luxQED to those of NNPDF3.0QED and
LUXqed17.
In all cases we will use the NNLO PDF sets, though the photon PDF depends only
mildly on the perturbative order (see Fig.~\ref{fig:photoncomp2}).
PDF uncertainties for the NNPDF sets are defined as the 68\% confidence level
interval and the central value as the midpoint of this range.
This is particularly relevant for NNPDF3.0QED for which, due to
non-Gaussianity in the replica distribution, PDF errors computed as
standard deviations can differ by up to one order of magnitude as
compared to the 68\% CL intervals.

\subsection{Drell-Yan production}
\label{sec:inclusiveWZproduction}

We begin by examining the role of PI contributions in neutral-current Drell-Yan
production.
We will study this process in three different kinematic regions of the
outgoing lepton pair: around the $Z$ peak, at low invariant masses,
and at high invariant masses.
We start with the $Z$ peak region, defined as $60 \le M_{ll} \le 120$
GeV, where $M_{ll}$ is the lepton-pair invariant mass, and focus
on the  central rapidity region $|y_{ll}|\le 2.5$, relevant for
ATLAS and CMS.\footnote{We have verified that similar results
  hold for the forward rapidity region, $2.0 \le y_{ll} \le 4.5$,
relevant for LHCb.}
This region provides the bulk of the Drell-Yan measurements included
in modern PDF fits and therefore assessing the impact of PI
contributions is particularly important here.

\begin{figure}[t]
\begin{center}
  \includegraphics[scale=0.38]{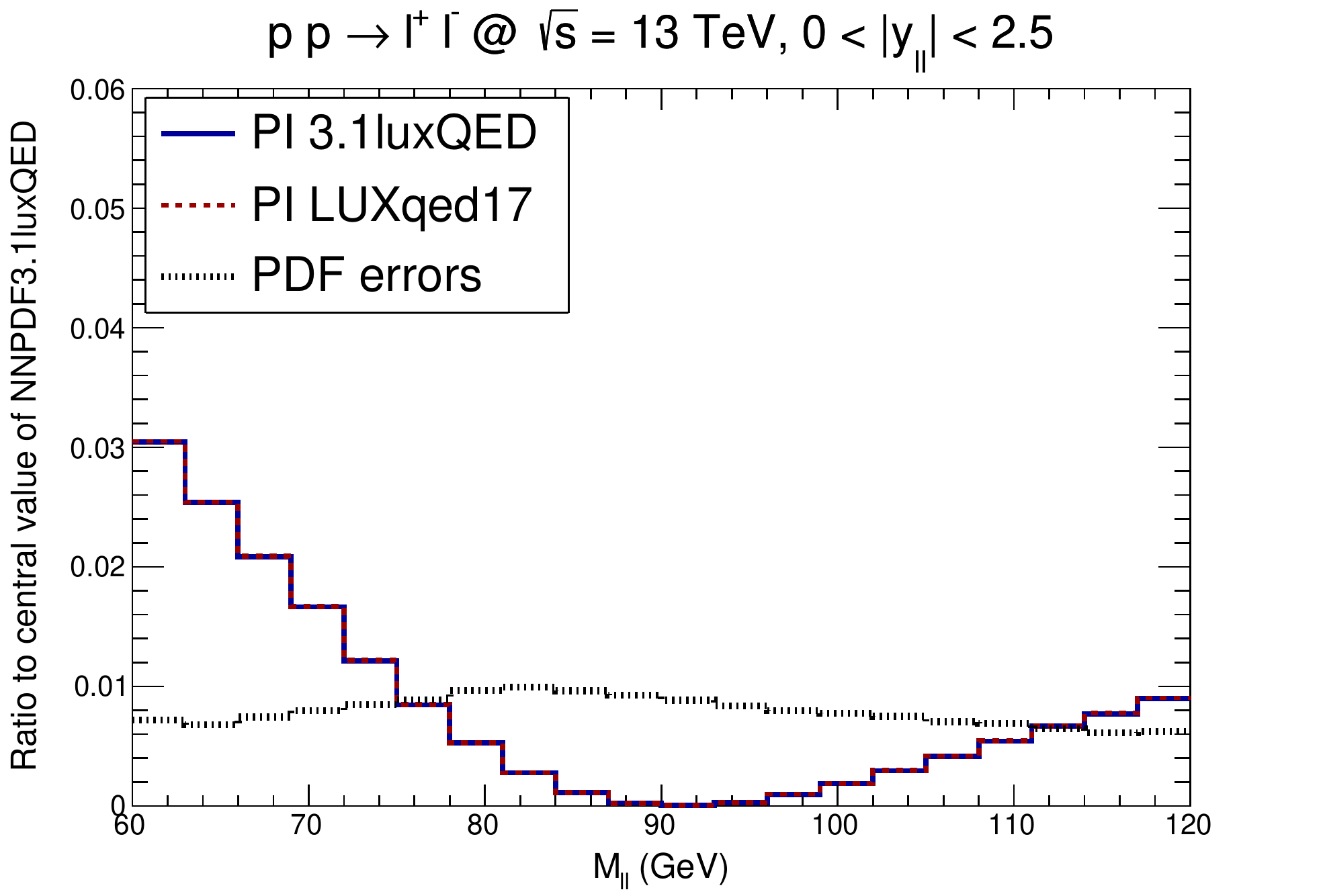}
  \includegraphics[scale=0.38]{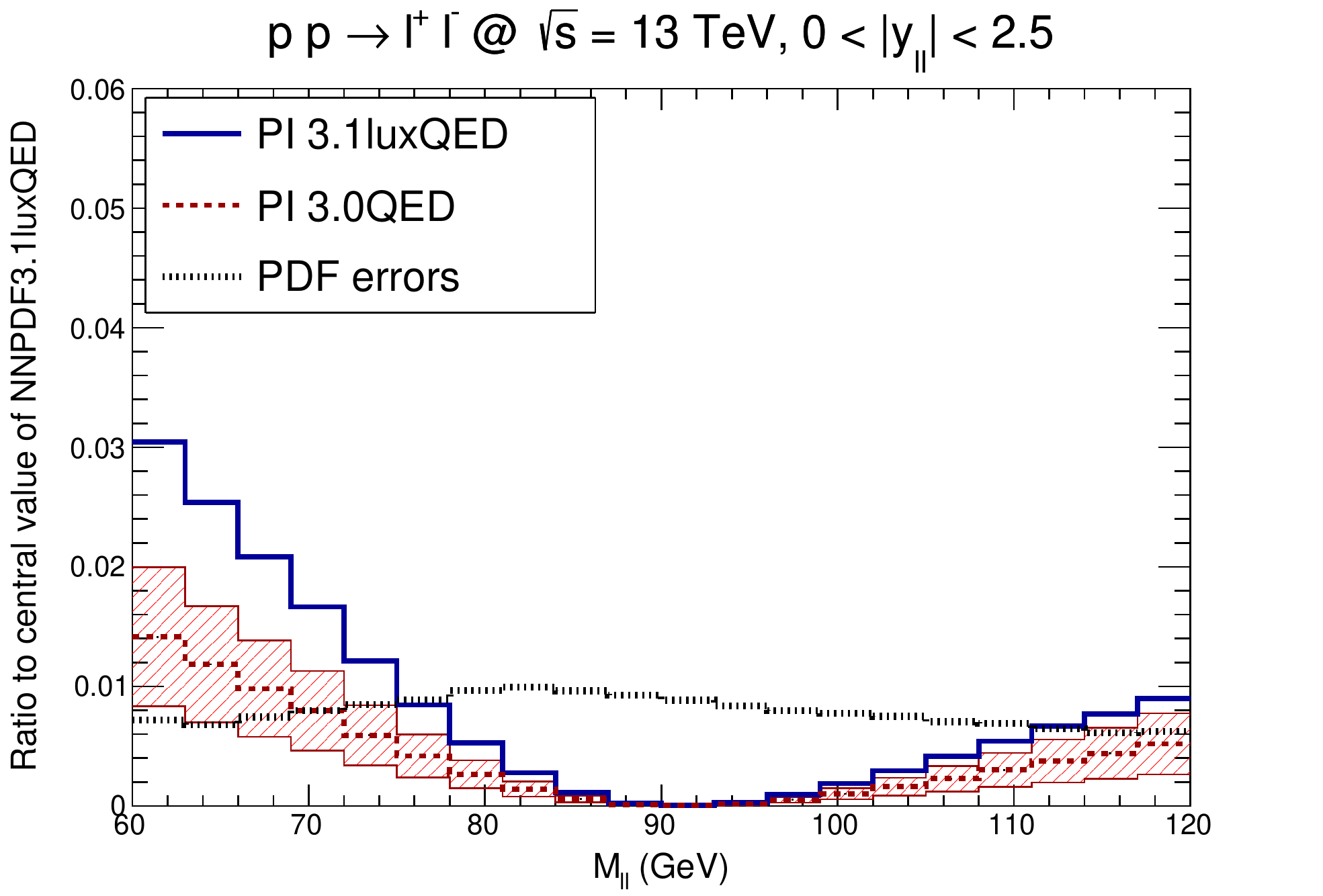}
  \caption{\small The ratio of photon-initiated
  contributions to the corresponding quark- and gluon-initiated ones
  for neutral current Drell-Yan production as function of the lepton-pair
  invariant mass $M_{ll}$ in the $Z$ peak region and central rapidities
  $|y_{ll}|\le 2.5$ at $\sqrt{s}=13$ TeV.
  We compare NNPDF3.0QED, LUXqed17, and NNPDF3.1luxQED, with the PI
  contributions in each case normalized to the central value of the latter.
  The NNPDF3.0QED uncertainty band is represented by the red band.
  For reference, we also indicate the value of the PDF uncertainties in
  NNPDF3.1luxQED\@.\label{fig:phenoDYcentral}
   }
\end{center}
\end{figure}

In Fig.~\ref{fig:phenoDYcentral} we show the ratio of the PI
contributions to the corresponding quark- and gluon-initiated
contributions for Drell-Yan production as a function of $M_{ll}$ at
$\sqrt{s}=13$ TeV in the $Z$ peak region.
We compare the predictions of NNPDF3.0QED, LUXqed17, and
NNPDF3.1luxQED, with the PI contributions normalised to the central
value of NNPDF3.1luxQED.
For reference we also show the value of the PDF uncertainties in
NNPDF3.1luxQED.

We find that PI effects for this process are at the permille level for
$M_{ll}\sim M_Z$ but they become larger as we move away from the $Z$
peak, reaching 3\% for $M_{ll}=60$ GeV. At the lower edge of the
$M_{ll}$ region the contribution of the PI channel exceeds the level
of PDF uncertainty, highlighting the sensitivity of this distribution
to the photon PDF.
We find that NNPDF3.1luxQED and LUXqed17 lead to a larger PI
contribution as compared to NNPDF3.0QED at low $M_{ll}$.
As the PI contribution is only significant away from the $Z$-peak,
where the bulk of the cross-section lies, these effects may be
reasonably neglected in the integrated cross-sections.

Fig.~\ref{fig:phenoDYcentral} demonstrates that the PI contributions
in NNPDF3.1luxQED and LUXqed17 lead to very similar results for
Drell-Yan production around the $Z$ peak.
We have verified that this similarity holds also
for the low and high mass kinematic regions, as well as for the rest
of processes studied in this section.
In the following discussion we will therefore restrict ourselves to
comparisons between NNPDF3.0 and NNPDF3.1luxQED.

We now move to study the low- and high-mass regions, defined as
$15 \le M_{ll} \le 60$ GeV and $M_{ll}\ge 400$ GeV
respectively.
Drell-Yan low-mass measurements have been presented by ATLAS, CMS, and
LHCb~\cite{Aad:2014qja,CMS:2014jea,Chiapolini:1528259}, with the two-fold
motivation of providing input for PDF fits and to study QCD in complementary
kinematic regimes.
The high-mass region is relevant for BSM searches that exploit
lepton-pair final states, such as those expected in the presences of
new heavy gauge bosons $W'$ or
$Z'$~\cite{Aaboud:2017buh,Khachatryan:2016zqb}.

\begin{figure}[t]
  \begin{center}
    \includegraphics[scale=0.38]{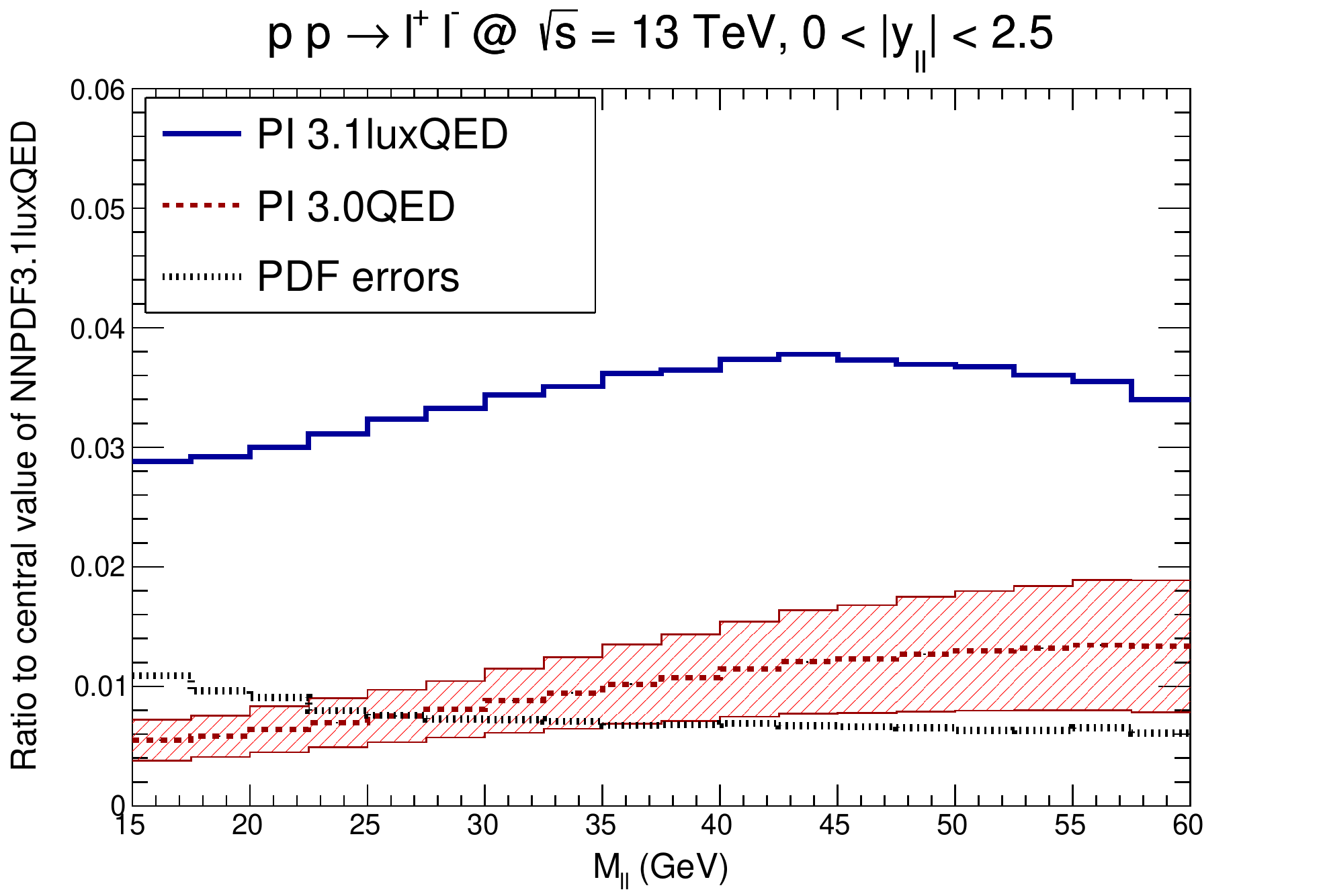}
    \includegraphics[scale=0.38]{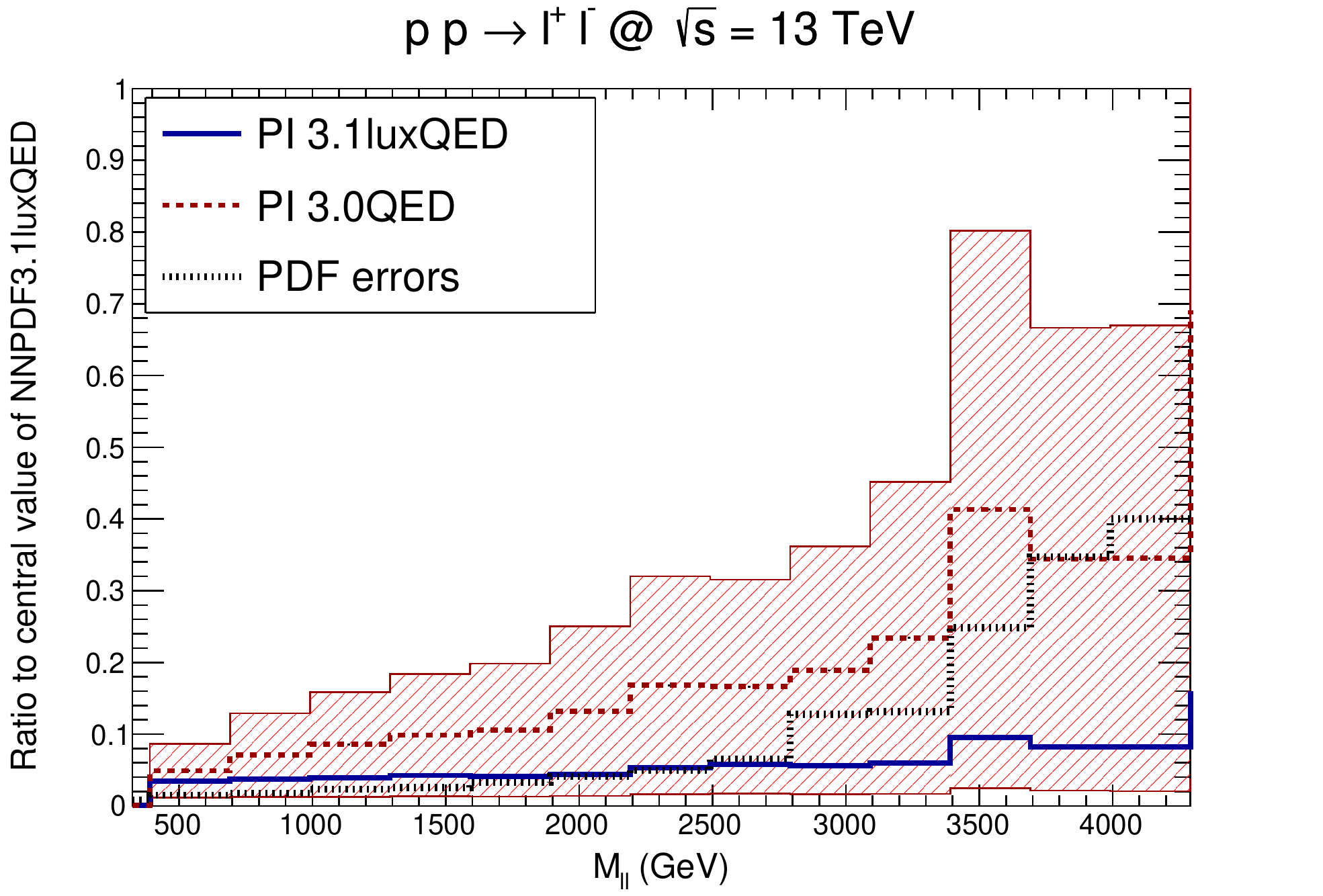}
    \caption{\small Similar representation as the right panel of Fig.~\ref{fig:phenoDYcentral} for the low (left)
      and high (right plot) invariant
    mass regions, defined as 15 GeV $\le M_{ll} \le$ 60 GeV and $M_{ll}\ge 400$ GeV respectively. Please note that the right figure is plotted in a larger $y$-axis range in comparison to previous plots.
    \label{fig:phenoDYlowmass}
    \label{fig:phenoDY1}
  }
\end{center}
\end{figure}

In Fig.~\ref{fig:phenoDYlowmass} we show the same comparison as in the
right panel of Fig.~\ref{fig:phenoDYcentral} for the low- and
high-mass regions.
In the low-mass case, the PI effects are more significant than in the $Z$-peak
region, being between 3\% and 4\% for most of the $M_{ll}$ range, consistently
larger that the PDF uncertainty.
We find that PI effects in
NNPDF3.1luxQED can be up to a factor three larger than in the
NNPDF3.0QED case due to the corresponding differences in the photon
PDF at small $x$.
In the case of the high-mass region, we observe that the effect
of the PI contribution computed with NNPDF3.1luxQED is comparable to
the PDF uncertainties for $M_{ll}\gsim 3$ TeV, eventually becoming
as large as $\simeq 10\%$ of the QCD cross-section.
These effects are markedly smaller than in NNPDF3.0QED, where shifts
in the cross-section up to $\simeq 80\%$ due to PI contributions were
allowed within uncertainties.

To conclude this discussion on Drell-Yan at the LHC, we have evaluated
the ratio of the LO PI contributions to the NLO QCD cross-sections for
the kinematics of the ATLAS high-mass Drell-Yan measurements at 8
TeV~\cite{Aad:2016zzw}.
Both the Bayesian reweighting study of the ATLAS
paper~\cite{Aad:2016zzw} and the analysis of Ref.~\cite{Giuli:2017oii}
indicate that this dataset has a considerable sensitivity to PI
contributions if NNPDF3.0QED is used as a prior.
Here we revisit this process to assess how the picture changes when
using NNPDF3.1luxQED.

\begin{figure}[t]
\begin{center}
  \includegraphics[scale=0.38]{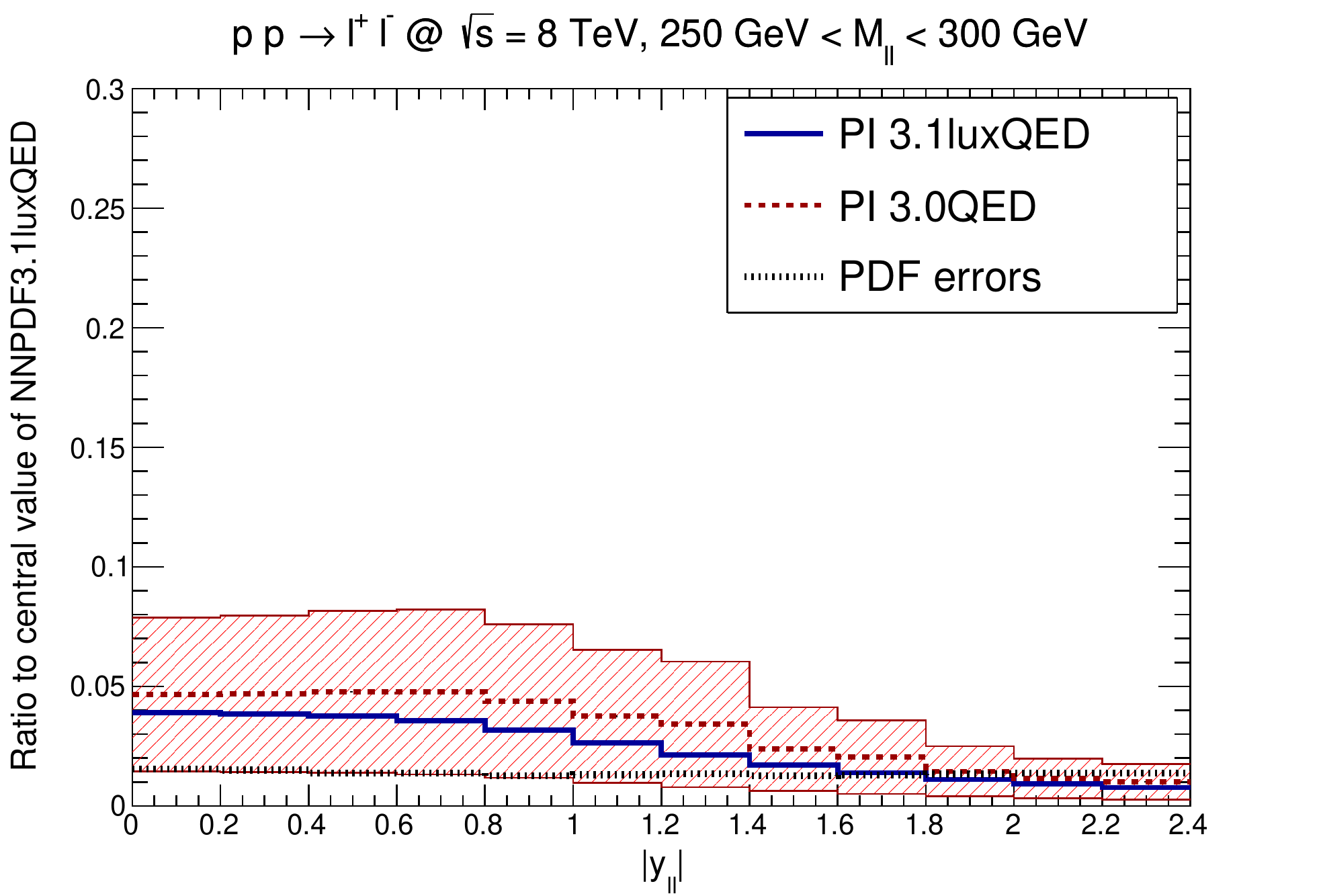}
  \includegraphics[scale=0.38]{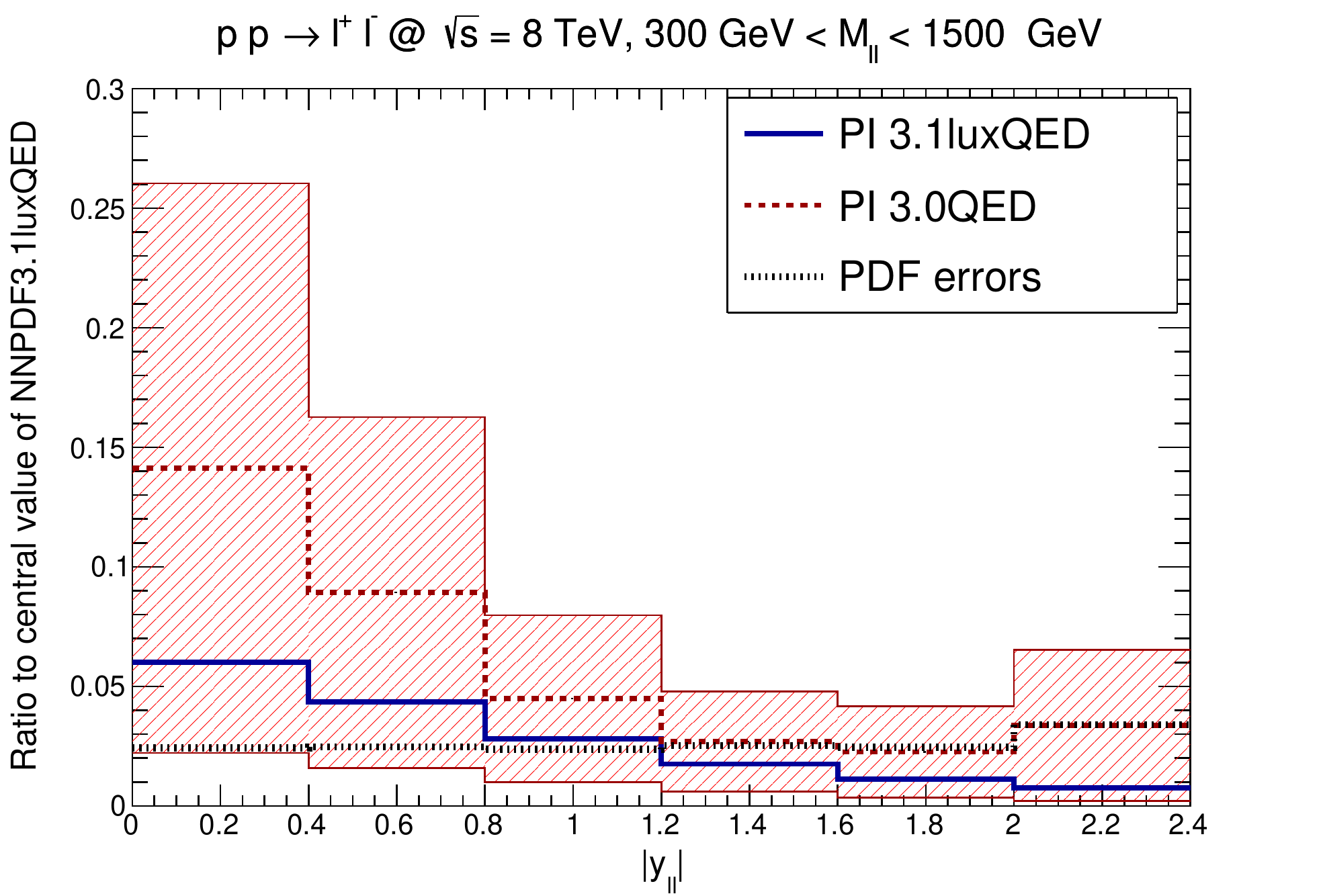}
  \caption{\small
Same as Fig.~\ref{fig:phenoDY1} for the kinematics of the ATLAS high-mass DY
measurement at 8 TeV.
We show results for the lepton-pair rapidity distributions $|y_{ll}|$ in two
different invariant mass bins, $250$ GeV$\le M_{ll} \le 300$ GeV (upper) and
$300$ GeV$\le M_{ll} \le 1500$ GeV (lower plots).
    \label{fig:phenoHMDY}
  }
\end{center}
\end{figure}

In Fig.~\ref{fig:phenoHMDY} we show the ratio of PI over QCD
contributions for the lepton-pair rapidity distributions $|y_{ll}|$ in
Drell-Yan at 8 TeV for two invariant mass bins, $250$
GeV$\le |M_{ll}| \le 300$ GeV and $300$ GeV$\le |M_{ll}| \le 1500$
GeV.
As can be seen, with NNPDF3.0QED the effects of the PI contribution at
large invariant masses can be as large as 25\% of the QCD
cross-section.
This shift is larger than the corresponding experimental uncertainties, which
are typically at the percent level, explaining the sensitivity of NNPDF3.0QED 
to this dataset.
From Fig.~\ref{fig:phenoHMDY} we observe that the PI contribution
becomes smaller when using NNPDF3.1luxQED, though its effect is still
comparable to the experimental uncertainties.
Indeed, the PI contribution in the highest invariant mass bin ranges
between 6\% of the QCD cross-section in the central rapidity region and 1\% for
$|y_{ll}|=2.5$.
This comparison confirms that the PI contribution is important for the
quantitative description of the Drell-Yan process above the $Z$ peak.

\subsection{Vector-boson pair production}
\label{sec:inclusiveVVproduction}

The production of vector-boson pairs is of particular interest for the LHC
physics program.
Firstly, they probe the electroweak sector of the SM and provide bounds on
possible anomalous couplings.
Secondly, this final state appears in several BSM scenarios and
therefore many searches involve the detection of pairs of weak vector
bosons.
When the invariant mass of the vector-boson pair $m_{VV}$ is large,
the PI contributions, arising already at the Born level (see
Fig.~\ref{fig:photoninitiated}), are known to be
significant~\cite{Bierweiler:2012kw}.
Here we will examine the case of opposite-sign $W^+W^-$ production at
the LHC 13 TeV.

\begin{figure}[t]
\begin{center}
  \includegraphics[scale=0.38]{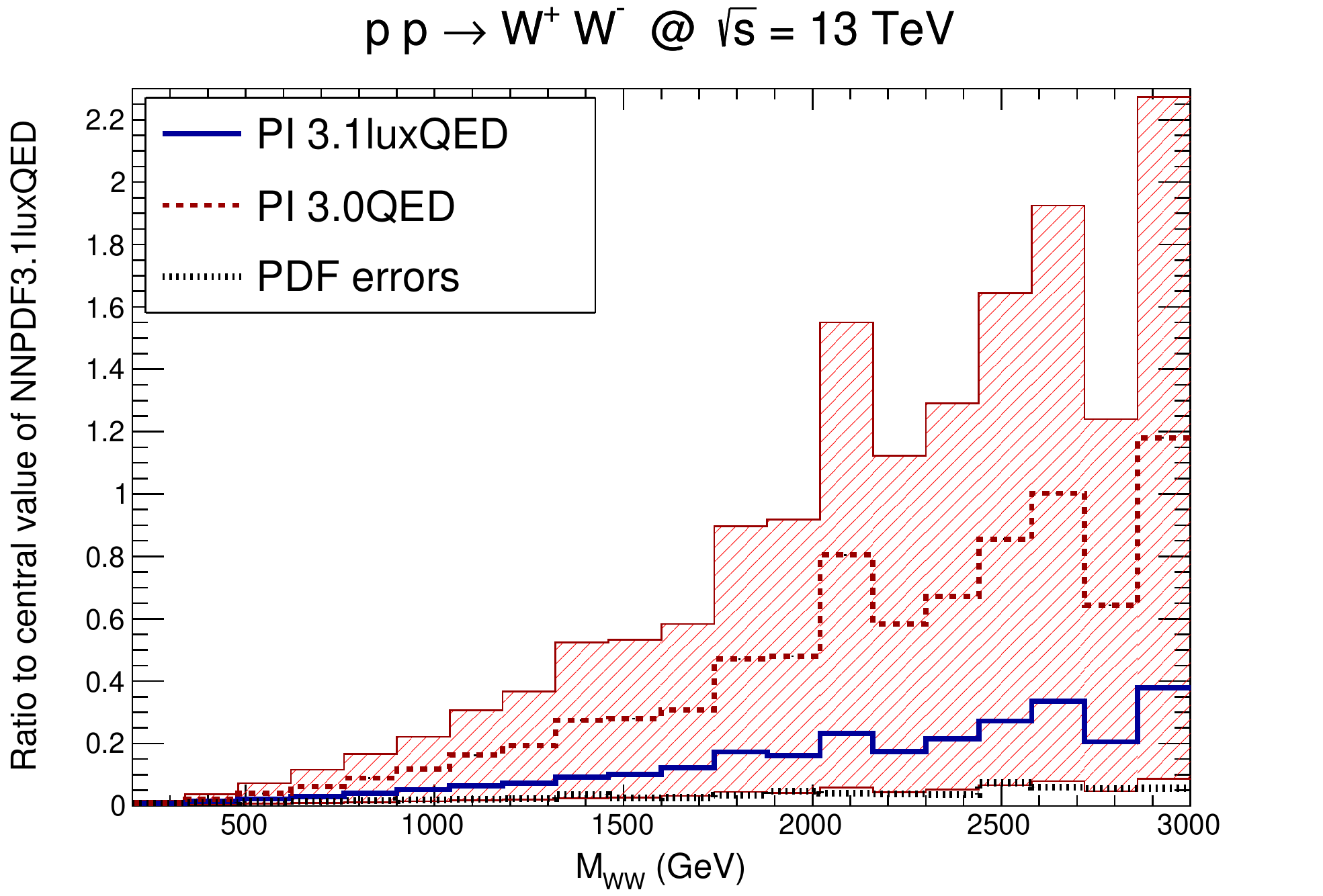}
  \includegraphics[scale=0.38]{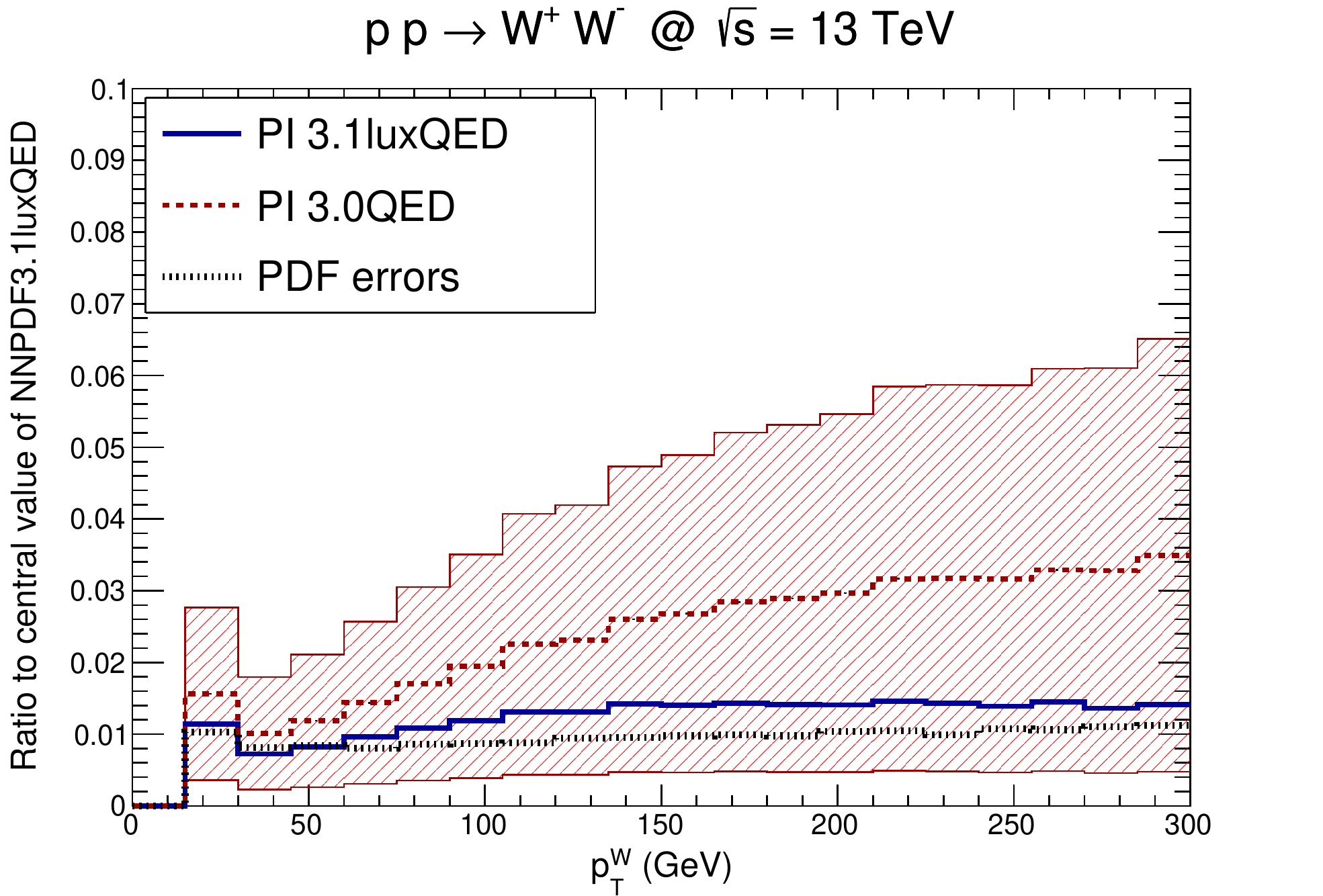}
  \caption{\small Same as Fig.~\ref{fig:phenoDY1} 
    for the production of a $W^+W^-$ pair, specifically
    for invariant mass distribution  $m_{WW}$ (left) and the
    transverse momentum of $W$ bosons $p_T^W$ (right plot).
    \label{fig:phenoWW1}
  }
\end{center}
\end{figure}

To assess the size of the PI contribution to this process, in
Fig.~\ref{fig:phenoWW1} we show the invariant mass distribution
$m_{WW}$ and the transverse momentum distributions of the $W$ bosons
$p_T^W$.
The $W$ bosons are taken to be stable and required to be in the central rapidity
region, $|\eta_W|\le 2.5$.
From the comparisons shown in Fig.~\ref{fig:phenoWW1} we find that for
the $m_{WW}$ distribution the PI contribution is larger than the PDF
uncertainties over the entire range considered.
In particular, when using NNPDF3.1luxQED, the size of the PI
contribution with respect to the total cross-section increases from
1\% at $M_{WW}\simeq 300$ GeV up to 35\% at $M_{WW}\simeq 3$ TeV.
As expected, the trend is similar using NNPDF3.0QED but with much
larger uncertainties. An upwards shift of the cross-section by a
factor two or larger would be allowed within PDF uncertainties
in this case.

The picture is rather different for the case of the transverse
momentum distribution of the individual $W$ bosons $p_T^W$.
Here we find that the PI contribution using NNPDF3.1luxQED are small,
around the $\simeq 1\%$ level, over the entire range in $p_T^W$
considered. Additionally, the effect of using NNPDF3.0QED instead is
not so dramatic, with an increase in the cross-section of a few
percent at most.
The differences between the two distributions arise from the fact that
the PI contribution to $W$ boson pair production is kinematically
enhanced only in the large $m_{WW}$ limit, irrespective of the value
of $p_T^W$.
The results of Fig.~\ref{fig:phenoWW1} suggest that current
measurements of this process from ATLAS and
CMS~\cite{Aaboud:2017qkn,Khachatryan:2015sga} might already be
sensitive to the photon PDF.

\subsection{Top-quark pair differential distributions}

\begin{figure}[t]
\begin{center}
  \includegraphics[scale=0.37]{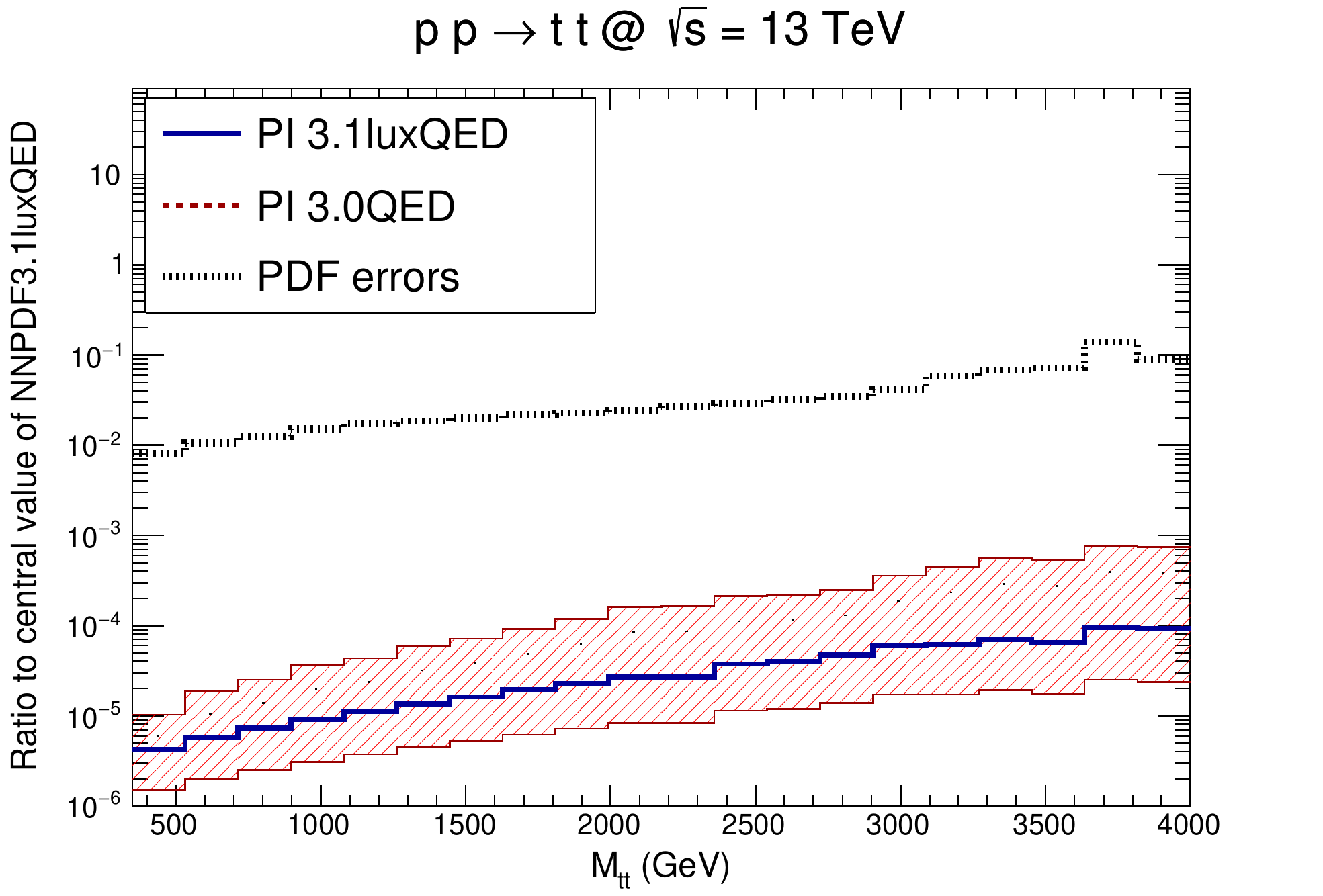}
  \includegraphics[scale=0.37]{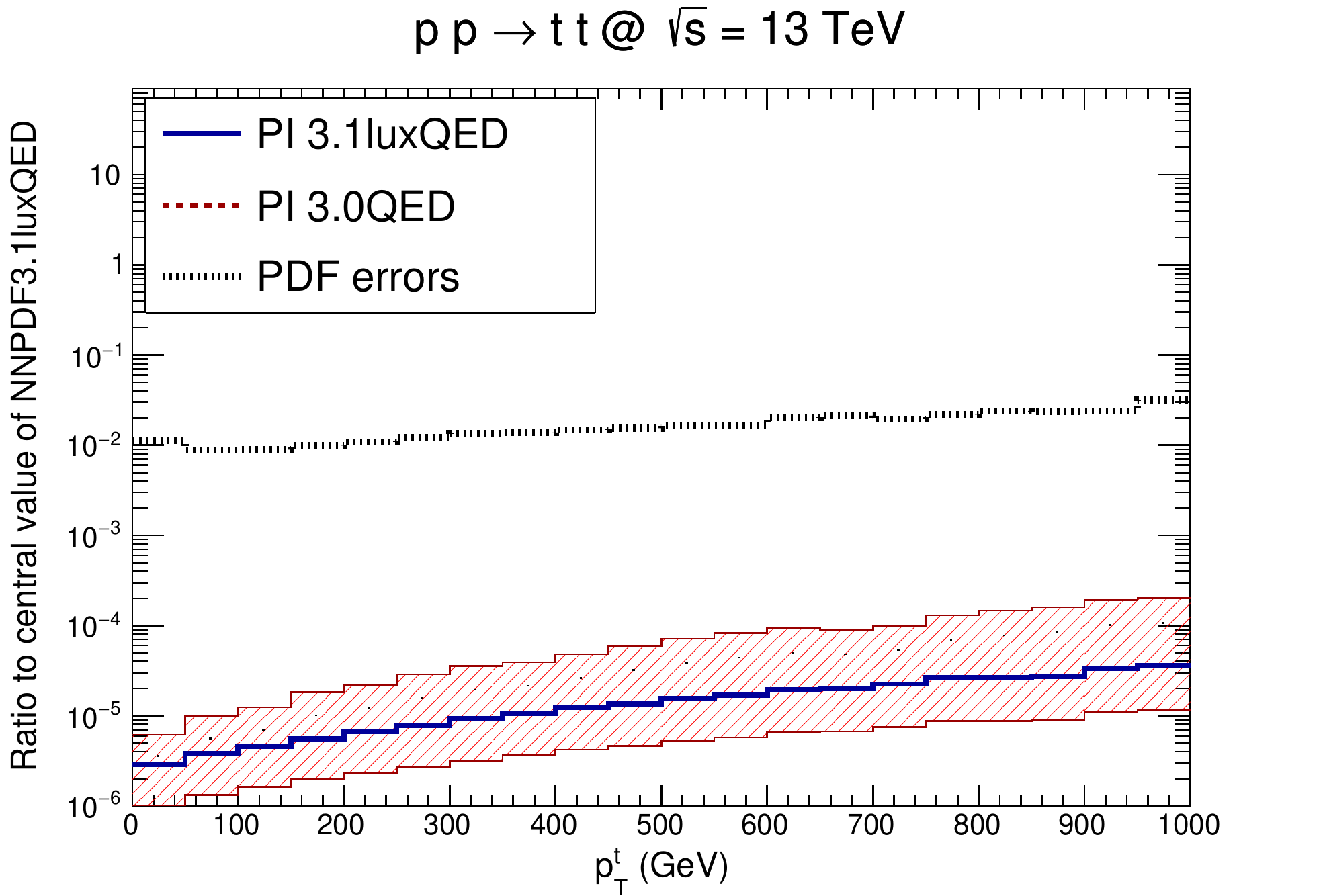}
  \caption{\small Same as Fig.~\ref{fig:phenoDY1}
  for the invariant mass distribution of top quark pairs $m_{t\bar{t}}$ (left)
  and the transverse momentum of top quarks $p_T^t$ (right plot) in top-quark
  pair production at 13 TeV.
    \label{fig:phenott1}
  }
\end{center}
\end{figure}

Next we turn to study the impact of the PI contributions on
differential distributions in top-quark pair production (see also
Refs.~\cite{Czakon:2017wor,Pagani:2016caq}). The {\tt APPLgrid} tables
generated for this process include only the $\gamma \gamma \rightarrow
t \bar{t}$ channel.
In Fig.~\ref{fig:phenott1} we show the ratio of the PI contribution
over the QCD cross-section for the invariant mass distribution of
top-quark pairs $m_{t\bar{t}}$ and the transverse momentum of the
single top quarks $p_T^t$ at the 13 TeV LHC.
Unlike in the case of high-mass Drell-Yan production, we find that the PI
contribution to top-quark pair production is negligible even for the highest
values of $m_{tt}$ and $p_T^t$ accessible at the LHC.
Indeed, in the case of NNPDF3.1luxQED the size of the PI contribution
is at the permille level at most.
Therefore, in theoretical calculations of top-quark pair production
with electroweak corrections, the PI contribution can be safely
neglected.

From the comparisons in Fig.~\ref{fig:phenott1} we also see that the
PI correction is somewhat larger in NNPDF3.0QED but with larger
associated uncertainties.
Even in this case, at the highest invariant masses the upper edge of the 68\% CL
interval indicates that corrections due to the PI contribution are at most
0.1\%.
We have also verified that the PI contribution to $t\bar{t}$ production is
phenomenologically negligible also for other distributions such as the
rapidity distribution of top quarks and top-antitop pairs, $y_t$ and
$y_{t\bar{t}}$, respectively.

\subsection{Higgs production in association with a vector bosons}

The last process that we consider in this section is Higgs production
in association with a vector boson, see
Fig.~\ref{fig:photoninitiated}.
PI corrections to this process are known to be significant. In the
Higgs Cross Section Working Group prediction for the total
cross-sections, where NNPDF3.0QED is used, the uncertainty due to the
PI contribution is the dominant source of theory
error~\cite{deFlorian:2016spz}.
To investigate this, we have generated and then combined exclusive
samples for $pp \to hW^+$ and $pp\to hW^+j$ with and without the PI
contribution.
Both the Higgs boson and the $W^+$ boson are required to be in the central
rapidity region, $|y_W|\le 2.5$ and $|y_h|\le 2.5$. No other kinematic cuts are
applied.

\begin{figure}[t] \begin{center}
    \includegraphics[scale=0.38]{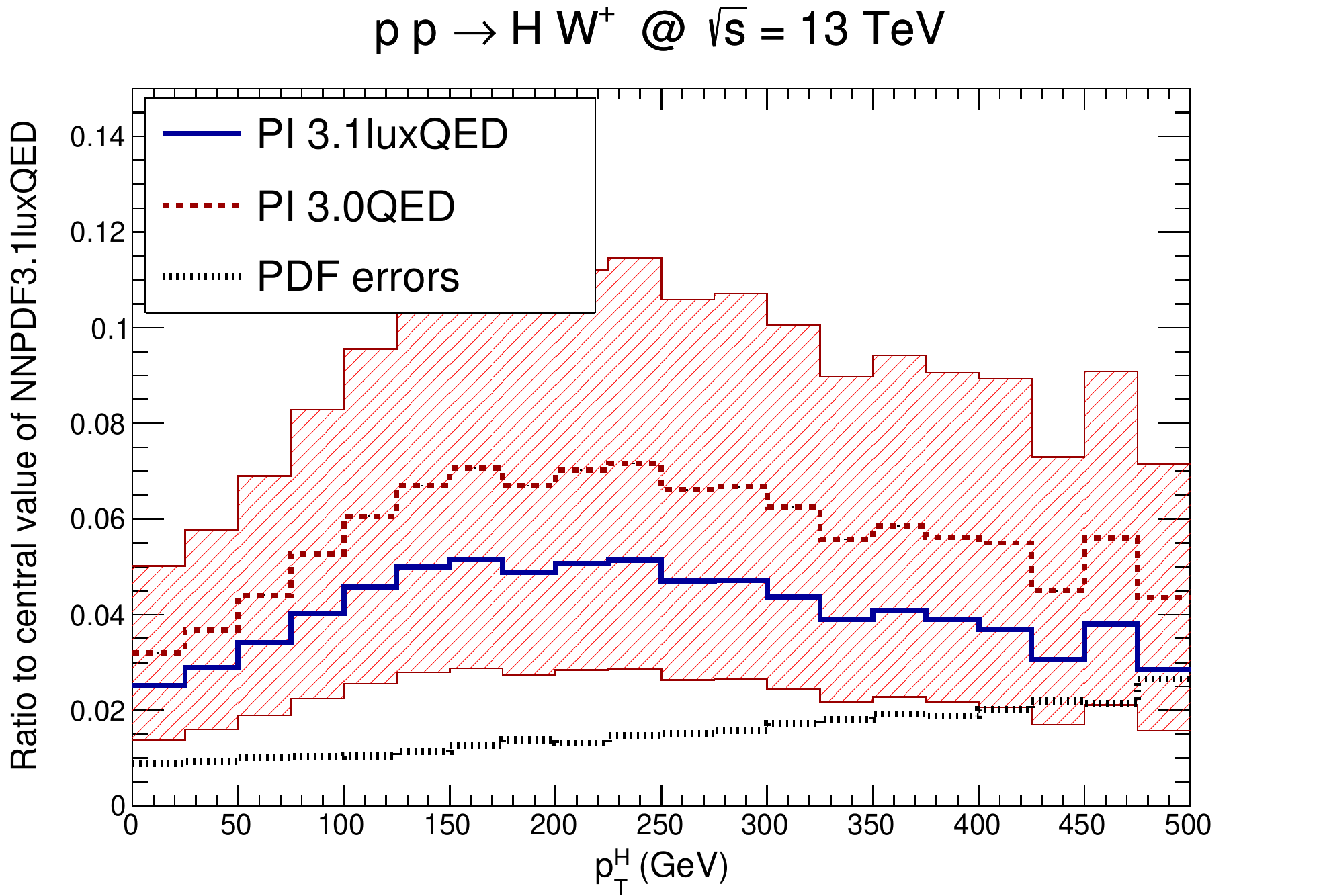}
    \includegraphics[scale=0.38]{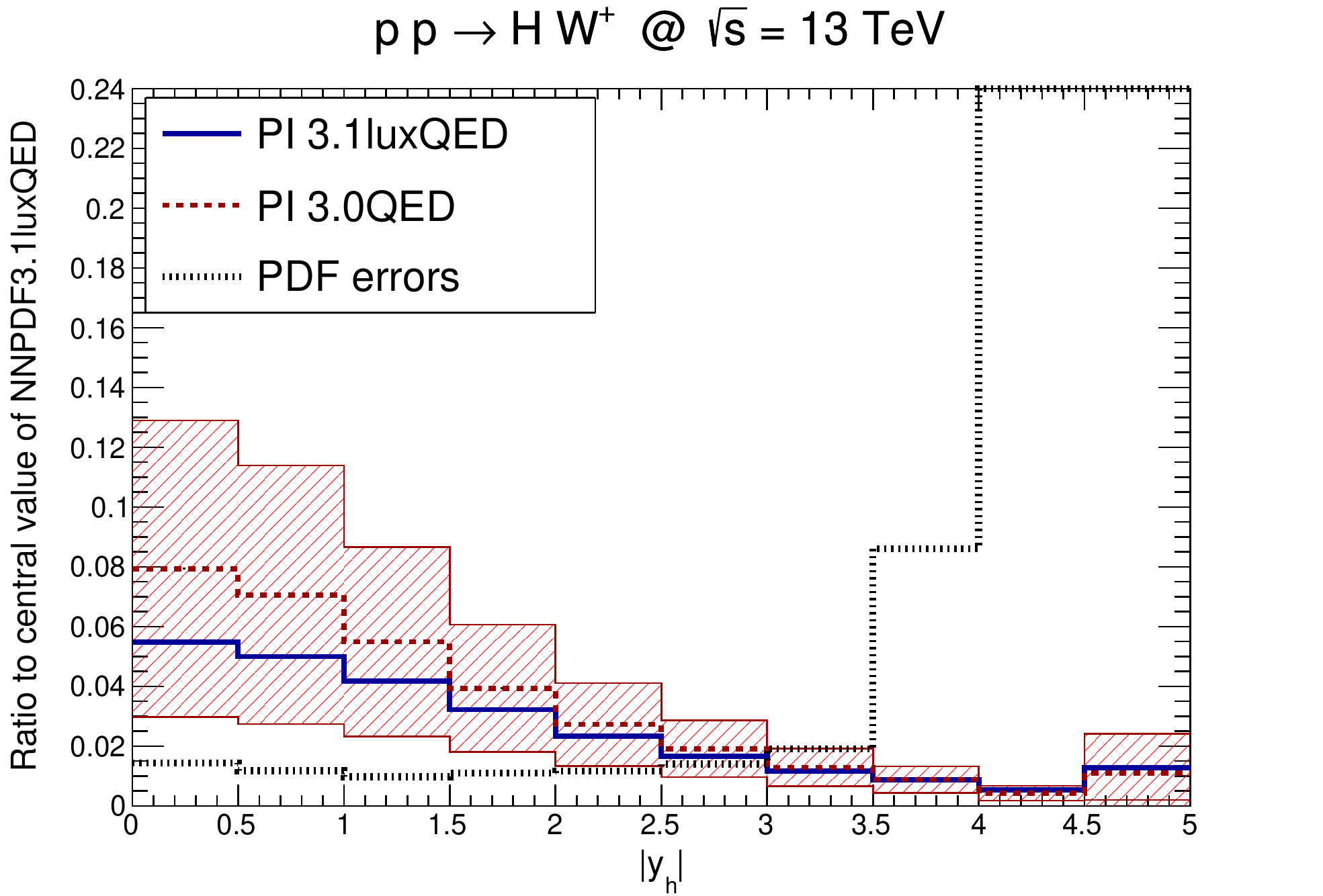}
    \caption{\small Same as Fig.~\ref{fig:phenoDY1} for Higgs production in
        association with a $W$ boson, for the  Higgs transverse momentum $p_T^h$
        distribution (left), and its rapidity $y_h$ distribution (right plot).
        \label{fig:phenoHW1} } 
\end{center} \end{figure}

In Fig.~\ref{fig:phenoHW1} we show the same comparison as in
Fig.~\ref{fig:phenoDYcentral} for the Higgs transverse momentum
$p_T^h$ and rapidity $y_h$ distributions.
In the case of the $p_T^h$ distribution, we find that PI effects can be up to
$5\%$ when using NNPDF3.1luxQED, with the largest effects localised at
intermediate values of $p_T^h\simeq 200$ GeV.
We also note that the shift induced by the PI contribution is bigger
than the PDF uncertainties.
Concerning the $y_h$ rapidity distribution, the PI contribution can be $\simeq
6\%$ in the central rapidity region when using NNPDF3.1luxQED, while it becomes
smaller as one moves to the forward region.

The comparisons of Fig.~\ref{fig:phenoHW1} illustrate that PI
contributions are relevant also for Higgs boson physics, including the
measurements of its couplings and branching fractions.

\section{Summary}
\label{sec:conclusion}

Parton distributions with QED effects and a photon
PDF are an essential
component in high-precision calculations of many LHC processes.
Previous NNPDF QED sets adopted a data-driven strategy to determine
the photon PDF, independently parametrising $\gamma(x,Q_0)$ and then fitting
it using constraints from Drell-Yan measurements at the LHC.
While this strategy minimised the theoretical bias due to model
assumptions, the lack of a precise experimental handle to constrain
the photon PDF led to large uncertainties.

With the development of the LUXqed framework, it is now
possible to constrain the photon PDF in terms of the accurately known
inclusive structure functions in lepton-hadron scattering.
In this work we have presented the
NNPDF3.1luxQED set, where the photon content of
the proton is determined by means of a
global PDF analysis supplemented by the LUXqed theoretical
constraint.
As a result, the uncertainty upon the photon PDF is considerably
reduced as compared to our previous NNPDF3.0QED determination,
down now to the level of a few percent.
We find that photons
carry up to 0.5\% of the total momentum of the proton,
and that the overall impact of the various types of QED effects
included in NNPDF3.1luxQED induce small but non-negligible
modifications in the quark and gluon PDFs.

We have then presented a first exploration of
the implications of NNPDF3.1luxQED 
for photon-initiated processes at the LHC.
We determine that the impact of PI contributions is consistent within uncertainties
with respect to previous estimates based on NNPDF3.0QED except for
the low-mass region $Q<M_Z$, and that
they  can be significant for many processes.
For instance, we find
corrections up to $\simeq 10\%$ for high-mass Drell-Yan and
up to $\simeq 20\%$  for $W^+W^-$ production.
In many cases, PI processes can be either comparable with or larger
than PDF uncertainties.
The uncertainty associated with these PI effects is in itself at the
level of a  few percent, so their overall effect is a shift of the cross-sections
as compared to the QCD-only calculation.

The NNPDF3.1luxQED set represents a state-of-the-art determination of the
PDFs of the proton including its photon component, accounting for
all relevant theoretical and experimental constraints.
This set is
therefore well suited for precision calculations of LHC processes.
The NNPDF3.1luxQED sets are available via the {\sc\small LHAPDF6}
interface~\cite{Buckley:2014ana}:
\begin{center}
\tt NNPDF31\_nlo\_as\_0118\_luxqed \\
\tt NNPDF31\_nnlo\_as\_0118\_luxqed \\
\end{center}
while the {\tt FiatLux} library, an open-source {\tt C++}
implementation of the LUXqed formalism, can be obtained from:
\begin{center}
  \url{https://github.com/scarrazza/fiatlux}
\end{center}
together with the corresponding documentation.

\bigskip
\begin{center}
\rule{5cm}{.1pt}
\end{center}
\bigskip

\subsection*{Acknowledgments}

We are grateful to our colleagues in the NNPDF Collaboration for many
illuminating discussions on the photon PDF and fits with QED corrections,
and specially to Stefano Forte for comments on this manuscript.
We are grateful to Aneesh Manohar, Paolo Nason, Gavin Salam, and Giulia
Zanderighi for discussions about the LUXqed formalism.
J.~R. is grateful to Lucian Harland-Lang for discussions about PDF fits with QED
effects.

V.~B., N.~H., and J.~R. are supported by an European Research Council Starting
Grant ``PDF4BSM''.
J.~R. is also partially supported by the Netherlands Organization for Scientific
Research (NWO).
S.~C. is supported by the HICCUP ERC Consolidator grant (614577) and by the
European Research Council under the European Union's Horizon 2020 research and
innovation Programme (grant agreement n$^{\circ}$ 740006).


\appendix
\section{NNPDF3.1luxQED fit quality}
\label{sec:appendix}

In this Appendix we collect the results of the fit quality
in the NNPDF3.1luxQED analysis and compare it to
those from its QCD-only counterpart, NNPDF3.1.
As customary in NNPDF analyses, these
$\chi^2$ values are computed using the experimental
definition of the covariance matrix, while
the $t_0$ definition~\cite{Ball:2009qv} was instead used during the fits
in order to avoid the D'Agostini bias.

In Table~\ref{tab:chi2} we list the values of $\chi^2/N_{\rm dat}$ from the NNPDF3.1 
  and NNPDF3.1luxQED NNLO fits for all the experiments included
  in the global analysis.
  We see that at the total dataset level, the fit quality
  is essentially identical in the two cases,
  yielding $\chi^2/N_{\rm dat}=1.148$.
  Also at the level of individual experiments there is
  good agreement between the two sets, with some small
  differences that are consistent with statistical fluctuations.
  Recall that these two fits are statistically independent, and
  therefore one expects fluctuations of the order $\Delta\chi^2
  \sim \sqrt{N_{\rm dat}}$
  in the values of individual experiments.

\begin{table}[tp]
\centering
  \small
  \renewcommand{\arraystretch}{1.13}
   \begin{tabular}{l C{2.4cm}C{3.0cm}}
     & \multicolumn{2}{c}{$\chi^2/N_{\rm dat}$}  \\
     & NNPDF3.1   & NNPDF3.1luxQED  \\
\toprule
NMC     &       1.30    &     1.31       \\
SLAC      &      0.75    &    0.71        \\
BCDMS      &     1.21    &    1.21        \\
CHORUS     &     1.11    &    1.11        \\
NuTeV dimuon     &   0.82      &  0.75          \\
\midrule
HERA I+II incl.  &      1.16    &   1.16         \\
HERA $\sigma_c^{\rm NC}$     & 1.45        &    1.48        \\
HERA $F_2^b$                &   1.11      &    1.11        \\
\midrule
DY E866 $\sigma^d_{\rm DY}/\sigma^p_{\rm DY}$     &  0.41        &  0.39          \\
DY E886 $\sigma^p$                              &  1.43       &   1.43         \\
DY E605  $\sigma^p$                             &  1.21       &   1.20         \\
CDF $Z$ rap                                     & 1.48       &    1.48        \\
CDF Run II $k_t$ jets                           &  0.87       &   0.88         \\
D0 $Z$ rap                                      &  0.60      &    0.60        \\
D0 $W\to e\nu$  asy                             &  2.70      &    2.68        \\
D0 $W\to \mu\nu$  asy                           &  1.56      &    1.57        \\
\midrule
ATLAS total                                 &  1.09        &    1.07        \\
ATLAS $W,Z$ 7 TeV 2010                      &  0.96        &    0.96        \\
ATLAS HM DY 7 TeV                           &  1.54        &    1.57        \\
ATLAS low-mass DY 7 TeV                      &  0.90        &  0.88          \\
ATLAS $W,Z$ 7 TeV 2011                      &  2.14        &   2.18         \\
ATLAS jets 2010 7 TeV                       &  0.94        &   0.91         \\
ATLAS jets 2.76 TeV                         &  1.03        &   1.02         \\
ATLAS jets 2011 7 TeV                       &  1.07        &   1.06         \\
ATLAS $Z$ $p_T$ 8 TeV $(p_T^{ll},M_{ll})$    &   0.93       &    0.93        \\
ATLAS $Z$ $p_T$ 8 TeV $(p_T^{ll},y_{ll})$    &   0.94       &    0.91        \\
ATLAS $\sigma_{tt}^{tot}$                    &   0.86      &     0.89       \\
ATLAS $t\bar{t}$ rap                        &   1.45     &     1.22       \\
\midrule
CMS total                               &   1.06       &  1.05          \\
CMS W asy 840 pb                   &  0.78        &   0.78         \\
CMS W asy 4.7 fb                 &   1.75       &   1.74         \\
CMS Drell-Yan 2D 2011                   &  1.23        &  1.27          \\
CMS W rap 8 TeV                 &  1.01        &      1.00      \\
CMS jets 7 TeV 2011                     &   0.84       &   0.81         \\
CMS jets 2.76 TeV                       &   1.03       &   1.02         \\
CMS $Z$ $p_T$ 8 TeV $(p_T^{ll},y_{ll})$  &  1.32        &   1.32         \\
CMS $\sigma_{tt}^{tot}$                  &  0.20       &   0.16         \\
CMS $t\bar{t}$ rap                      &  0.94      &   0.99         \\
\midrule
LHCb total                               &   1.47       &      1.47      \\
LHCb $Z$ 940 pb                             &   1.49       &    1.48        \\
LHCb $Z\to ee$ 2 fb                         &   1.14       &    1.11        \\
LHCb $W,Z\to \mu$ 7 TeV                     &   1.76       &    1.79        \\
LHCb $W,Z\to \mu$ 8 TeV                    &   1.37       &    1.36        \\
\bottomrule
{\bf Total }    &    \bf 1.148    &\bf  1.148   \\
\end{tabular}
\vspace{0.5cm}
\caption{\small The values of $\chi^2/N_{\rm dat}$ from the NNPDF3.1 
  and NNPDF3.1luxQED NNLO fits for all the experiments included
  in the global analysis.
  These $\chi^2/N_{\rm dat}$ values have
  been computed using the experimental definition.
}
\label{tab:chi2}
\end{table}

From the comparison shown in Table~\ref{tab:chi2} we can conclude
that while QED effects lead to small differences at the quark
and gluon PDF level (see Fig.~\ref{fig:qcd_nnpdf31qed-vs-nnpdf31}),
the fit quality is still very similar to the QCD-only fit.
This implies that these small QED effects can be reabsorbed into the
PDFs, leading to the same quantitative description of the datasets
included in the analysis.
Note that the situation is likely to be rather different once measurements
directly sensitive to the photon content of the proton, such
as the ATLAS high-mass Drell-Yan at 8 TeV, are
included into the fit.

\clearpage

\providecommand{\href}[2]{#2}\begingroup\raggedright\endgroup

\end{document}